\documentclass[fleqn,usenatbib]{rasti}

\usepackage{newtxtext,newtxmath}
\usepackage[T1]{fontenc}

\DeclareRobustCommand{\VAN}[3]{#2}
\let\VANthebibliography\thebibliography
\def\thebibliography{\DeclareRobustCommand{\VAN}[3]{##3}\VANthebibliography}

\usepackage{graphicx}	% Including figure files
\graphicspath{ {plots/} }
\usepackage{amsmath}	% Advanced maths commands
\usepackage{hyperref}
\usepackage{multirow}
\usepackage{longtable}
\usepackage{xcolor}
\usepackage{newtxtext,newtxmath}
\usepackage{subcaption}

\title[Timing With NICER And Deadtime]{The Effects of Instrumental Deadtime on NICER Timing Products}

\author[R. Webbe et al.]{
Robbie Webbe,$^{1,2}$\thanks{\url{https://orcid.org/0000-0003-1689-3723}}
and A. J. Young$^{1}$\thanks{\url{https://orcid.org/0000-0003-3626-9151}}
\\
$^{1}$H. H. Wills Physics Laboratory, Tyndall Avenue, Bristol, BS8 1TL, UK \\
$^{2}$IRAP, Université de Toulouse, CNRS, CNES, 9 Avenue du Colonel Roche,31028 Toulouse, France
}

\date{Accepted XXX. Received YYY; in original form ZZZ}

\pubyear{2023}

\begin{document}
\label{firstpage}
\pagerange{\pageref{firstpage}--\pageref{lastpage}}
\maketitle

\begin{abstract}
The X-ray Timing Instrument as part of the Neutron Star Interior Composition Explorer has the potential to examine the time-domain properties of compact objects in regimes not explored by previous timing instruments, due to its combination of high effective area and timing resolution. We consider the effects of instrumental deadtime at a range of effective countrates in a series of observations of the X-ray binary GX 339-4 to determine what effect deadtime has on photometric and Fourier frequency-domain products. We find that there are no significant inconsistencies across the functional detectors in the instrument, and that in the regimes where instrumental deadtime is a limiting factor on observations that previous approaches to dealing with deadtime, as applied to RXTE and other detectors, are still appropriate, and that performing deadtime corrections to lightcurves before creating Fourier products are not necessary at the count rates considered in our analysis. 
\end{abstract}

% Include between one and six keywords.

%%%%%%%%%%%%%%%%%%%%%%%%%%%%%%%%%%%%%%%%%%%%%%%%%%

%%%%%%%%%%%%%%%%% BODY OF PAPER %%%%%%%%%%%%%%%%%%

\section{Introduction}
\label{sec:intro}

The Neutron Star Interior Composition Explorer payload currently in operation on the International Space Station (ISS) contains the X-ray Timing Instrument (XTI), which is a successor to the Rossi X-ray Timing Explorer (\emph{RXTE}) mission. The XTI operates by concentrating incoming X-ray emission from a $30\text{ arcmin}^2$ field of view. While it was initially designed for detailed study of Neutron Stars \citep{arzoumanian_neutron_2014}, the high timing resolution also makes it ideal for the study of other fast-variable X-ray sources like X-Ray Binaries (XRBs).

\begin{figure}
    \includegraphics[width=\columnwidth,trim={0 0.25cm 0.25cm 0},clip]{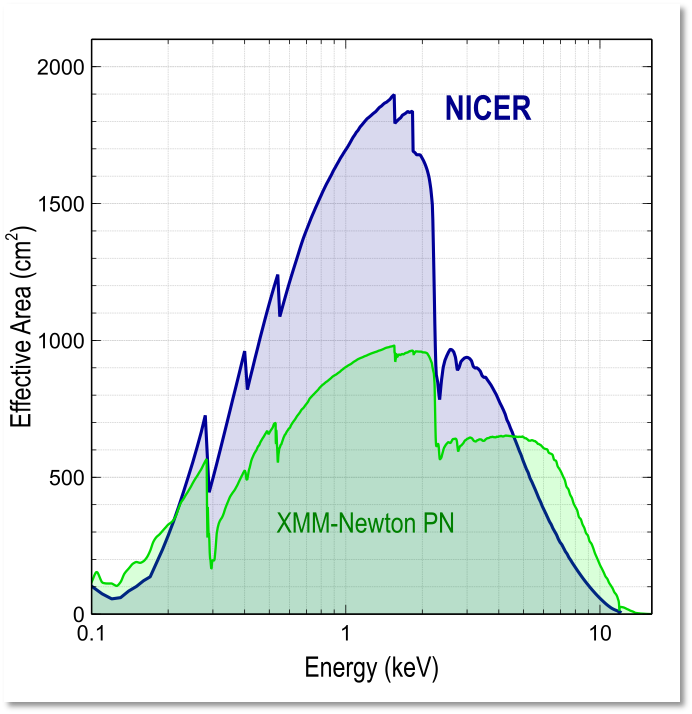}
    \caption{Effective area for the NICER X-ray Timing Instrument in comparison with the PN detector onboard XMM-Newton. Figure reproduced from \protect\cite{arzoumanian_neutron_2014}.}
    \label{fig:effarea}
\end{figure}

The XTI primarily operates in the energy range from 0.2--12.0\,keV, thus overlapping with previous X-ray timing missions like \emph{RXTE}. It has a higher spectral resolution than other X-ray timing missions, of 85\,eV at 1\,keV and 137\,eV at 6\,keV, making it comparable to other observatories like \emph{Chandra} and \emph{XMM-Newton}, albeit with a far larger effective area. At soft X-ray energies the XTI has an effective area of $\sim 1793 \text{ cm}^2$ \citep{okajima_performance_2016} which is $\sim2\times$ greater than that of \emph{XMM-Newton}. See Fig. \ref{fig:effarea} which is reproduced from \cite{arzoumanian_neutron_2014}. While it has been designed specifically to handle high count rates, the main strength of the XTI for our purposes is, however, its timing resolution. In comparison with previous missions the timing resolution possible, of the order of $\sim300 \text{ ns}$ \citep{prigozhin_nicer_2016}, is at least two orders of magnitude greater than that achievable with \emph{XMM-Newton}, and approximately $\sim25\times$ better than that of \emph{RXTE}.

The XTI consists of a series of 56 concentrator modules \citep{okajima_performance_2016}, which concentrate, rather than focus, incoming high energy events on to a series of Focal Plane Modules (FPMs) at the rear of the instrument. Each of these FPMs contains one Silicon Drift Detector (SDD) responsible for recording the incoming event energy and arrival time. Although there are a total of 56 FPMs in the instrument, four of them have been inactive since launch. Signals from the FPMs are passed through two passbands: fast and slow \citep{prigozhin_nicer_2016}. These two passbands are optimised for spectral and timing performance separately. The `fast' passband provides the greater timing resolution, but has a higher read noise level, meaning it is not suitable for use with lower energy events. All events trigger a `slow' chain analysis with lower timing resolution but higher spectral resolution. This combination of chains allows for the combination of high spectral and timing data, and also forms part of the filtering of spurious events not coming from the source of interest. Alongside the creation of events from incident X-rays from the source of interest, the FPMs will also record events from: incident charged particles; the local high-energy background; radiation secondary to incident charged particle events. To properly screen for these we consider the ratio of the pulse heights from the `fast' and `slow' chains.

All events follow a common processing pipeline during which non-source events should be appropriately screened. The first step in the pipeline is to calibrate all events on a per-Measurement and Power Unit (MPU) basis. This step accounts for time biases and calculates the slow and fast-chain Pulse Invariant (\texttt{PI}) values which can then be interpreted as event energy. For \emph{NICER} \texttt{PI} values are recorded in units of 10\,eV, so a \texttt{PI} value for an event of 167 would correspond to an event energy of 1670\,eV. At this stage \texttt{PI\_RATIO} values are also calculated as the ratio 
\begin{equation}
\texttt{PI\_RATIO} = \frac{\texttt{PI}} {\texttt{PI\_FAST}}
\end{equation}
where \texttt{PI} is the pulse height from the slow chain, and \texttt{PI$\_$FAST} is the pulse height from the fast chain if present. At this stage we also perform the first screening step. Events which are flagged as being either undershoot, overshoot or forced trigger events, i.e., they cannot be calibrated, are usually removed at this step. Undershoot events are a natural response to the functioning of the SDDs and to optical loading, and occur when the detector discharges a cumulative build up of low level charge not related to incident X-ray events. Overshoot events occur when high energy events interact with the detectors at energies which are beyond the calibrated maximum. These are often related to incoming cosmic rays which deposit some energy while passing through the detector and as such are not related to X-ray events of interest. Forced trigger events occur when the detector is sampled by software when events are not incident. This forms part of the calibration of individual detectors but is also not related to X-ray events. The second step involves screening the exposure for times which are likely to be affected by contamination. At this stage we can filter for several variables including: elevation of the source above the Earth limb or bright Earth; times when the observing will be affected by the South Atlantic Anomaly; times when undershoot events are above a given rate (usually due to optical loading of the detectors); times when the overshoot events are above a given rate (times with high cosmic ray event rate). This allows us to create Good Time Intervals (GTIs) during which we can have an appropriate level of confidence in the calibrated event lists containing a high level of true X-ray events. The third stage of the pipeline collates the events from the seven MPUs into a single data file and allows for further screening by event types or event energies. At this stage we perform background screening using the \texttt{PI$\_$RATIO} values. Standard \emph{NICER} background screening occurs by means of a `trumpet filter', so named for the formula used to make cuts by \texttt{PI\_RATIO} values. Events with 
\begin{equation}
\texttt{PI\_RATIO} > fastconst + \frac{fastsig} {\texttt{PI}} + fastquart \times \texttt{PI}^3
\end{equation}
are filtered out as non X-ray events. Events which come from X-ray detections should have a \texttt{PI\_RATIO} of $\sim1.0$ \citep{prigozhin_nicer_2016}. In order to account for minor variation in $fastconst$ within a $10\%$ tolerance is therefore set at $1.1$. $fastsig$ is dependent on the noise level threshold in the fast chain and is set at 1200\,eV. $fastquart$, which was a measure of the quartic gain error in the fast chain, is currently set to zero due to improvements in the fast chain calibration following initial testing\footnote{\url{https://heasarc.gsfc.nasa.gov/lheasoft/ftools/headas/nicerclean.html}}. Finally, FPMs and MPUs are screened for unusual behaviour when compared with the ensemble of detectors. If an FPM or MPU is subject to unusual undershoot, overshoot or noise rates at greater than  $2\sigma$ from the average that individual detector can be excluded. Detectors which are suffering from telemetry saturation (see Section \ref{subsec:intro-deadtime} for further details) can also be screened at this time.

\subsection{Deadtime}
\label{subsec:intro-deadtime}
Deadtime, generally speaking, is used to describe any time during which the detector in question is unable to process the arrival of any new incoming events. For some X-ray instruments any adjustments for deadtime are made during the processing of the data. Such processing occurs automatically in the production of scientific products, for example, by means of the \texttt{epiclccorr} task in the \emph{XMM-Newton} processing pipeline, or as accounted for in deadtime corrected exposure times by \emph{SWIFT/XRT}. In the case of \emph{NICER} there are two principal sources of deadtime: deadtime produced by the detector reading out individual event data from each MPU; deadtime caused when the buffer for the FPM reaches its maximum capacity. For the purposes of this analysis we will be focusing on the effects of the first type of deadtime, which we will henceforth call `detector deadtime'. The effects of the second type of deadtime, which will henceforth be referred to as `telemetric deadtime', appear as a different phenomenon in the processing of the data. Telemetric deadtime presents as a series of fragmented Good Time Intervals (GTIs) which create gaps in what would otherwise be good sections of observation not limited by other effects like the South Atlantic Anomaly or elevation above the Earth limb. This phenomenon is described in the \emph{NICER} Data Analysis Guide\footnote{\url{https://heasarc.gsfc.nasa.gov/docs/nicer/data_analysis/nicer_analysis_tips.html}}, and occurs at countrates of the order $\sim20000$ and above, and we will not consider the effects of this deadtime in the remainder of this chapter. We can make this delineation between detector and telemetric deadtime due to the manner in which we will be approaching the creation of time-domain and Fourier-domain products. In all cases we will only be using continuous segments of lightcurves which have at least a minimum segment length, and as such will only be considering data during periods where telemetric deadtime is not an effect. We reject segments which may be contaminated, even by short breaks in the GTIs, in order to produce products which can be tested as robustly as possible for detector deadtime in isolation.

Detector deadtime is recorded on an event-by-event basis in the event files produced by \emph{NICER}, and on a second-by-second basis in the housekeeping files. \emph{NICER} offers very high timing resolution, allowing us to probe variability down to ms or lower scales with its timing resolution on scales of $\sim300 \,\text{ns}$, and with the expectation that we may need to probe variability on sub-second timescales we therefore base our work in the following sections upon the event-by-event deadtime information included in the event files, as it will allow us to consider sub-second variability with confidence that fluctuations in the deadtime caused by events over the duration of one second are being appropriately considered.

\bigskip
In this paper we aim to determine whether previous approaches to mitigating the effects of instrumental deadtime on time-domain and frequency-domain products are appropriate for the newer generations of timing instruments. We do this by means of the analysis of observational data from \emph{NICER}. In section \ref{sec:methods} we describe the observations analysed in this paper, and the manner in which time-domain and frequency-domain products are created. We report the effects of instrumental deadtime on these observations in section \ref{sec:results}. In section \ref{sec:discuss} we discuss the effects of this deadtime on time-domain and Fourier products which are widely used in the literature to assess \emph{NICER} observations, and compare these results with currently accepted practices. In section \ref{sec:concs} we consider the impact of these results for the future use of \emph{NICER} observational data.

\section{Methods and Data}
\label{sec:methods}
We propose to correct for the effects of detector deadtime in observational \emph{NICER} data by considering the total deadtime affecting lightcurve bins in time-domain products and then correcting the lightcurve for the loss of detecting time. In doing so we will be making the assumption that the count rate over the course of individual time bins is at a constant level, and that we can therefore extrapolate the counts observed while the detector is capable of detecting events at the same rate during any deadtime. We will make this correction by summing the deadtime for all events impacting individual time bins, determining this as a proportion of the total time bin, and then dividing by the proportion of the time bins which were available for observing.
\begin{equation}
        r_{\text{True}} = \frac{r_{\text{Obs}}}{1 - \tau_{\text{Dead,prop}}}
\end{equation}
where $r_{\text{Obs}}$ is the observed count rate in a given time bin, $\tau_{\text{Dead,prop}}$ is the proportion of that time bin lost to detector deadtime, and $r_{\text{True}}$ is the expected `true' count rate during this time bin. We determine the total deadtime in a given time bin by finding the sum for all time bins of:
\begin{itemize}
    \item Deadtime for events falling within the bin where the end of the event deadtime is still within the time bin.
    \item The start of any deadtime for an event which occurs during the bin time, but which rolls over in to the next time bin
    \item The remainder of the deadtime for any events which began before the start of the time bin, but where the deadtime continued in to the time bin in question
\end{itemize}

As such, 
\begin{equation}
        \tau_{\text{Dead,prop}} = \frac{\tau_\text{Tot}}{\Delta T \times N_{\text{Det}}}
\label{eq:dtcorr}
\end{equation}
where 
$\tau_\text{Tot}$ is the total deadtime from all events in the time bin, $\Delta T$ is the time binning for the lightcurve, and $N_{\text{Det}}$ is the number of detectors active during that time. In order to give as full an account as possible of the deadtime from all events we use the barycenter corrected, but uncalibrated event files for all detectors which are contributing to the lightcurve.

Time-domain analysis of X-ray sources is one way in which we may probe the geometries and evolution of compact objects. Analysis of lightcurves and their Fourier transforms has been key to the detection of periodic oscillations in X-ray Binaries \citep[for a review see][ and associated references]{ingram_review_2019}, but such measurements require a high degree of confidence in such time-domain and Fourier products.
Detector deadtime has the potential to significantly affect the form of lightcurves, and Fourier power spectra derived from them. This is particularly true at high count rates where the proportional deadtime could take up significant fractions of individual time bins. It is likely that this proportional effect would act to suppress the count rates when the true count rates are particularly high. Any effects on lightcurves would in turn affect power spectra and any cross-spectral products upon which these are based. One proposed approach which has been applied to NuSTAR is to use the cross-spectrum created by taking the Fourier transform of lightcurves from multiple detectors \citep{bachetti_no_2015,bachetti_no_2018}. For instruments with several detectors this approach is not necessarily practical, and so the currently accepted best practice for instruments with multiple detectors \citep[see ][etc.]{morgan_rxte_1997,nowak_rossi_1999} is to make a frequency-dependent correction to the power spectra dependent, as proposed by \cite{zhang_dead-time_1995} as an amendment to the approach developed by \cite{van_der_klis_fourier_1989}. In this approach the deadtime affected Poisson noise level to be subtracted from a normalised power spectrum is defined as

\begin{align}
    P_{\text{deadtime}}(f) =& \frac{2}{r_{\text{e}}}\ \left(\left[\ 1 \ - \ 2r_{\text{pe}}\ \tau_{\text{d}}\left(1\ -\ \frac{\tau_{\text{d}}}{2t_{\text{b}}}\right)\right] \right. \nonumber \\ 
    &-\  \frac{N_f \ - \ 1}{N_f} \ r_{\text{pe}}\ \tau_{\text{d}}\left(\ \frac{\tau_{\text{d}}}{t_{\text{b}}}\ \right)\cos(\ 2\pi t_{\text{b}} f\ )  \nonumber \\ 
    &\left. +\  r_{\text{pe}}\ r_{\text{vle}}\left[\frac{\sin(\pi \ \tau_{\text{vle}} \ f)}{\pi f}\right]^2 \ \right)
	\label{eq:zhang-noise}
\end{align}
where $t_{\text{b}}$ is the length of the time bins in the lightcurve being Fourier transformed, $N_f$ is the number of frequency bins in the power spectrum, $r_{\text{e}}$ is the average count rate during the observation, $r_{\text{pe}}$ is the average count rate per detector during the observation, $r_{\text{vle}}$ is the count rate for very large events (those with a significantly higher than usual deadtime), $\tau_{\text{d}}$ is the deadtime associated with regular events, and $\tau_{\text{vle}}$ is the characteristic deadtime for large events.

In sections \ref{subsec:dtperfpm} and \ref{subsec:dteffects} we will consider the effects of deadtime corrections to power spectra, and their associated lightcurves in the context of nine observations with \emph{NICER} of the X-ray Binary system GX 339-4. We examine three observations each with relatively low, medium or high count rates. We have selected observations of the same target in order to mitigate for any differences in background when looking at different sections of the sky, and choose observations taken during a similar period in the lifetime of the instrument to try and mitigate any changes in performance due to deterioration in the instrument over time. The details of these nine observations are listed in Table \ref{tab:obs}.
\begin{table*}
    \centering
    \caption{Details of nine observations of the X-ray Binary GX 339-4 used for the analysis of deadtime effects in Sections \ref{subsec:dtperfpm} and \ref{subsec:dteffects}. We list the observation date, total exposure time during the observation, and the total time during our GTIs, and the average countrate across all GTIs for the energy range 0.2--15.0\,keV.}
    \label{tab:obs}
    \begin{tabular}{lccccr}
    \hline
    Obs ID & Observation Start & Observation End & Exposure (s) & GTI (s) & Rate (/s) \\
    \hline
         3133010102 & 2021-01-21 00:23 & 2021-01-21 20:53 & 2332 & 1102 & 31.2 \\
         3133010105 & 2021-01-23 23:39 & 2021-01-24 21:41 & 3968 & 2353 & 52.2 \\
         4133010262 & 2021-10-12 05:27 & 2021-10-12 22:42 & 2336 & 865 & 73.8 \\
         3558010201 & 2021-02-14 01:05 & 2021-02-14 22:59 & 4569 & 3939 & 749 \\
         4133010103 & 2021-03-26 03:22 & 2021-03-26 23:49 & 6111 & 4445 & 3202 \\
         4133010104 & 2021-03-27 01:04 & 2021-03-27 23:07 & 8709 & 7605 &  3935 \\
         4133010127 & 2021-04-19 02:36 & 2021-04-19 23:05 & 11622 & 11529 & 6780 \\
         4133010151 & 2021-05-22 04:10 & 2021-05-22 23:05 & 2922 & 2922 & 6933 \\
         4133010144 & 2021-05-07 05:41 & 2021-05-07 22:57 & 1645 & 1645 & 7024 \\
         \hline
    \end{tabular}
\end{table*}

These nine observations are processed in accordance with the standard \emph{NICER} processing pipeline. We use default values for all variables relating to individual event screening, with the default energy range set to 0.2--15.0\,keV and to screen for undershoot, overshoot and forced trigger events. We do make slightly more conservative choices with the time-based screenings, in order to avoid as best as possible any exposure time which could be affected by telemetric deadtime. When creating the GTIs we therefore use the more restrictive values of: \texttt{elv} = 30$^\circ$, which restricts good times to when the distance to the Earth limb is at least 30$^\circ$; \texttt{br\_earth} = 40$^\circ$, which restricts good times to when the distance to the bright Earth is at least 40$^\circ$; \texttt{underonly\_range} = "0-50", restricts the maximum undershoot event rate to 50 ct/s or lower; \texttt{cor\_range} = "4.0-", which restricts the magnetic cut-off rigidity to 4.90\,GeV/c.

\section{Results}
\label{sec:results}

\subsection{Deadtime Across Detectors}
\label{subsec:dtperfpm}
\begin{figure*}
\includegraphics[width=\textwidth]{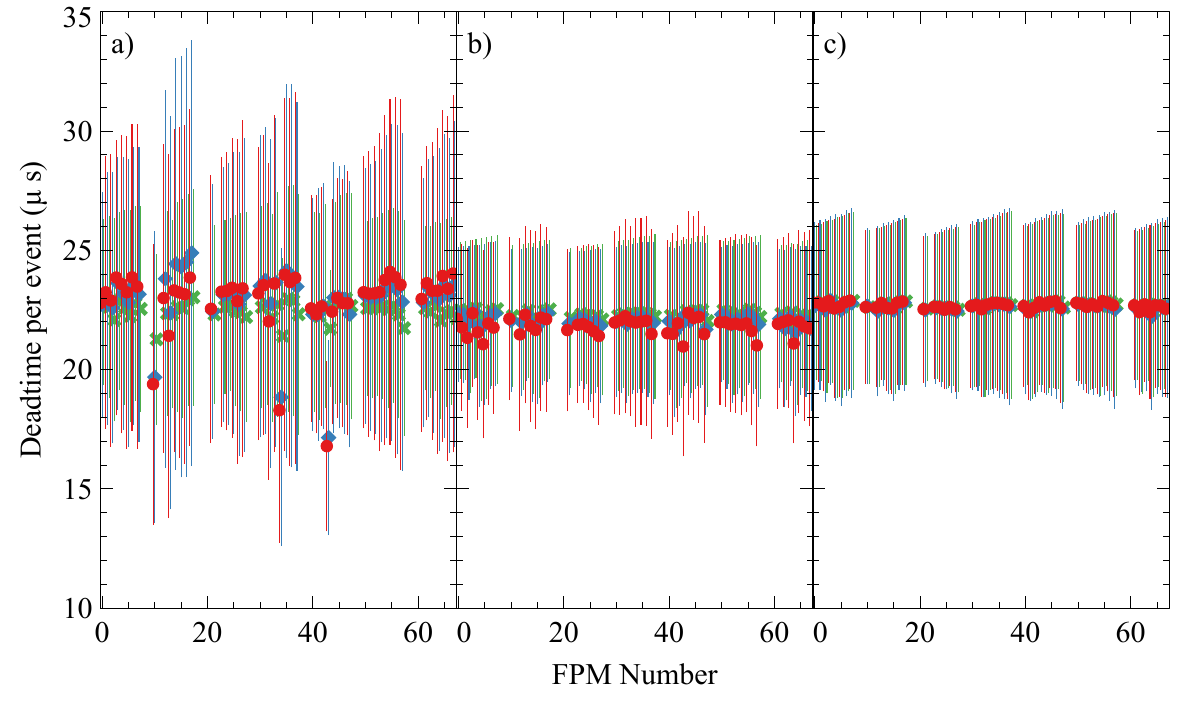}
\caption{Deadtime per recorded event across the different FPMs. Panels show the deadtime per event for a) low count rate (3133010102 red circles, 3133010105 blue squares, 4133010262 green crosses), b) medium count rate (3558010201 red circles, 4133010103 blue squares, 4133010104 green crosses), and c) high count rate observations (4133010127 red circles, 4133010144 blue squares, 4133010151 green crosses).}
\label{fig:dtperfpm}
\end{figure*}

We first consider whether there are any differences in the deadtime experienced by individual detectors. In Fig. \ref{fig:dtperfpm} we consider the average deadtime per event recorded by the individual FPMs in the cases where the average count rate is below 100 ct/s (observations 3133010102, 3133010105 and 4133010262), in the range from 100--4000 ct/s (observations 3558010201, 4133010103 and 4133010104), or above 4000 ct/s (observations 4133010127, 4133010144 and 4133010151).

In all nine observations we find that the average deadtime per recorded event is roughly constant, at $\sim$22$\mu$s per event, for all detectors. The only notable exceptions are for FPMs 10, 34 and 43 during observations 3133010102 and 3133010105 where the average deadtime per event is below $20\mu s$.

\begin{figure*}
\includegraphics[width=\textwidth]{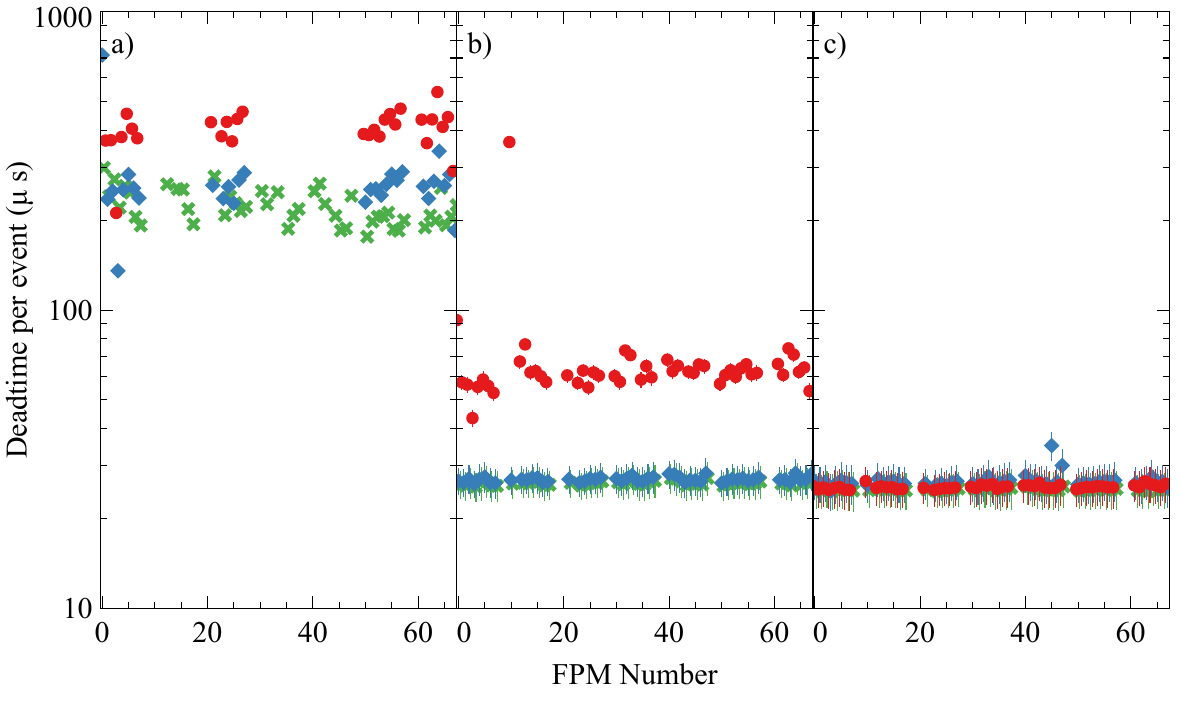}
\caption{Deadtime per event contributing to the lightcurve across the different FPMs. Panels show the deadtime per lightcurve event for a) low count rate (3133010102 red circles, 3133010105 blue squares, 4133010262 green crosses), b) medium count rate (3558010201 red circles, 4133010103 blue squares, 4133010104 green crosses), and c) high count rate observations (4133010127 red circles, 4133010144 blue squares, 4133010151 green crosses). In panel a) the error bars are smaller than the icons.}
\label{fig:dtperfpmlc}
\end{figure*}

In Fig. \ref{fig:dtperfpmlc} we consider whether there are any significant differences in the deadtime per event across the detectors as a function of the number of events which contribute to the final lightcurves, as this could display whether any detectors register high levels of events which are screened during the pipeline. We only consider the deadtime for events arriving during the good time intervals, as other events will be filtered when creating the GTIs.

The first result of note is that the average deadtime per lightcurve event tends to drop as the count rate increases. This would be expected if the number of screened events, and therefore the total amount of detector deadtime from screened events, was constant across a variety of observations. In this way, increasing the number of `good' counts would mitigate the effects of deadtime from screened events. Secondly, we also note that there are individual detectors in some of the low and medium count rate observations which show deviations from the average deadtime per lightcurve event. 
In the two earliest low count rate observations FPM 03 displays lower than average deadtime per X-ray event. In both observations the number of X-ray events are the same as for other detectors, while the deadtime is $\sim50\%$ less than that seen in other detectors. We also see a significantly higher deadtime per event for FPM 00 in observation 3133010105 where the number of lightcurve counts is approximately $\frac{1}{3}$ that of the other detectors, although this could be due to restrictions on the GTIs for that specific detector meaning it contributes to less of the duration of the lightcurve than other detectors.
The one anomaly from the medium count rate observations is detector 10 in observation 3558010201 appears to exhibit significantly more deadtime per X-ray event than any of the other active FPMs. During this observation, FPMs 11, 20, 22 and 60 were inactive, as they have been since launch, and FPMs 34 and 43, which have previously been identified as vulnerable to contamination by optical loading \citep[e.g.][]{Remillard2021,vivekanand_phase-resolved_2021} contain no counts which contribute to the lightcurve. Of the remaining 50 FPMs which contribute to the cleaned and screened lightcurve, detector 10 is a clear outlier, detecting $\sim28\%$ of the number of `good' events when compared to other detectors, while also experiencing a total deadtime $\sim66\%$ greater than the average for the other 49 contributing detectors. We show the full results in table \ref{tab:deadtime-fpm}.

\begin{table*}
	\centering
	\caption{Average and spread on deadtime experienced by different FPMs within observations. Values are presented as the minimum, mean and maximum average deadtime per recorded event during the duration of the observation, and average deadtime per X-ray event contributing to the lightcurve for that observation recorded during the GTIs of the observation.}
	\label{tab:deadtime-fpm}
	\begin{tabular}{lc|ccc|ccr}
            \hline
            \multicolumn{1}{c}{} &
            \multicolumn{1}{c}{} &
            \multicolumn{3}{c}{FPM Deadtime Per Recorded Event ($\mu s$)} &
            \multicolumn{3}{c}{FPM Deadtime Per X-ray Event ($\mu s$)} \\
		\hline
		OBSID & Average Rate (/s) & Minimum & Mean & Maximum & Min & Mean & Max \\
		\hline
            3133010102 & 31.2 & 16.80 & $22.95\pm1.33$ & 24.10 & 211.7 & $404.8\pm59.5$ & 538.3 \\
            3133010105 & 52.2 & 17.15 & $22.97\pm1.26$ & 24.89 & 135.4 & $270.6\pm92.9$ & 716.3 \\
            4133010262 & 73.8 & 21.27 & $22.49\pm0.37$ & 23.04 & 176.2 & $222.7\pm29.7$ & 300.0 \\
            3558010201 & 749.2 & 20.97 & $21.82\pm0.34$ & 22.38 & 43.42 & $68.33\pm43.53$ & 365.8 \\
            4133010103 & 3202.0 & 21.55 & $22.06\pm0.21$ & 22.43 & 26.17 & $27.07\pm0.54$ & 28.39 \\
            4133010104 & 3935.2 & 21.89 & $22.29\pm0.17$ & 22.56 & 25.60 & $26.33\pm0.43$ & 27.33 \\
            4133010127 & 6779.7 & 22.41 & $22.69\pm0.13$ & 22.89 & 24.92 & $25.53\pm0.40$ & 26.69 \\
            4133010151 & 6933.3 & 22.43 & $22.69\pm0.12$ & 22.88 & 24.63 & $25.02\pm0.24$ & 25.60 \\
            4133010144 & 7024.4 & 22.23 & $22.64\pm0.12$ & 22.90 & 24.70 & $26.64\pm1.40$ & 35.11 \\
		\hline
	\end{tabular}
\end{table*}

The general trend is such that the average deadtime per event recorded by the FPMs decreases with increasing X-ray count rates, both when considering the average deadtime per recorded event and when looking at the average deadtime per X-ray event during good time intervals. That these values tend to decrease with increasing X-ray count rate would indicate that deadtime can be affected by individual large deadtime events which are mitigated by the existence of large numbers of lower deadtime X-ray events when the count rate increases. If we examine the deadtimes of individual events across the observations, we find that the maximum deadtime associated with any one event is $79.1 \mu s$, with very variable numbers of such events being recorded during different observations. The two oldest observations, 3133010102 and 3133010105, record the greatest numbers of such high deadtime events with 128304 and 87426 in the two observations respectively. Observation 4133010103 recorded the smallest number of large deadtime events (818) despite having a total exposure time of 8ks which was double the length of three of the other observations. The relative rates of such events during observations is therefore very variable, with the greatest rate of high deadtime events being $11.51 ct/s$ during observation 3133010102, and the lowest being $0.10 ct/s$ during observation 4133010103. The number of such events which pass through calibration and screening to be categorised as X-ray events contributing to the lightcurves for these observations is also very variable. Only one of the largest deadtime events for observations 3133010102 and 3133010105 out of the total 128304 and 87426 recorded for those two observations respectively are passed through to the screened and calibrated event lists. By contrast 1871 of the 1946 recorded largest deadtime events for observation 4133010151 are X-ray events which form part of screened event list for the observation.

\begin{figure*}
\includegraphics[width=\textwidth]{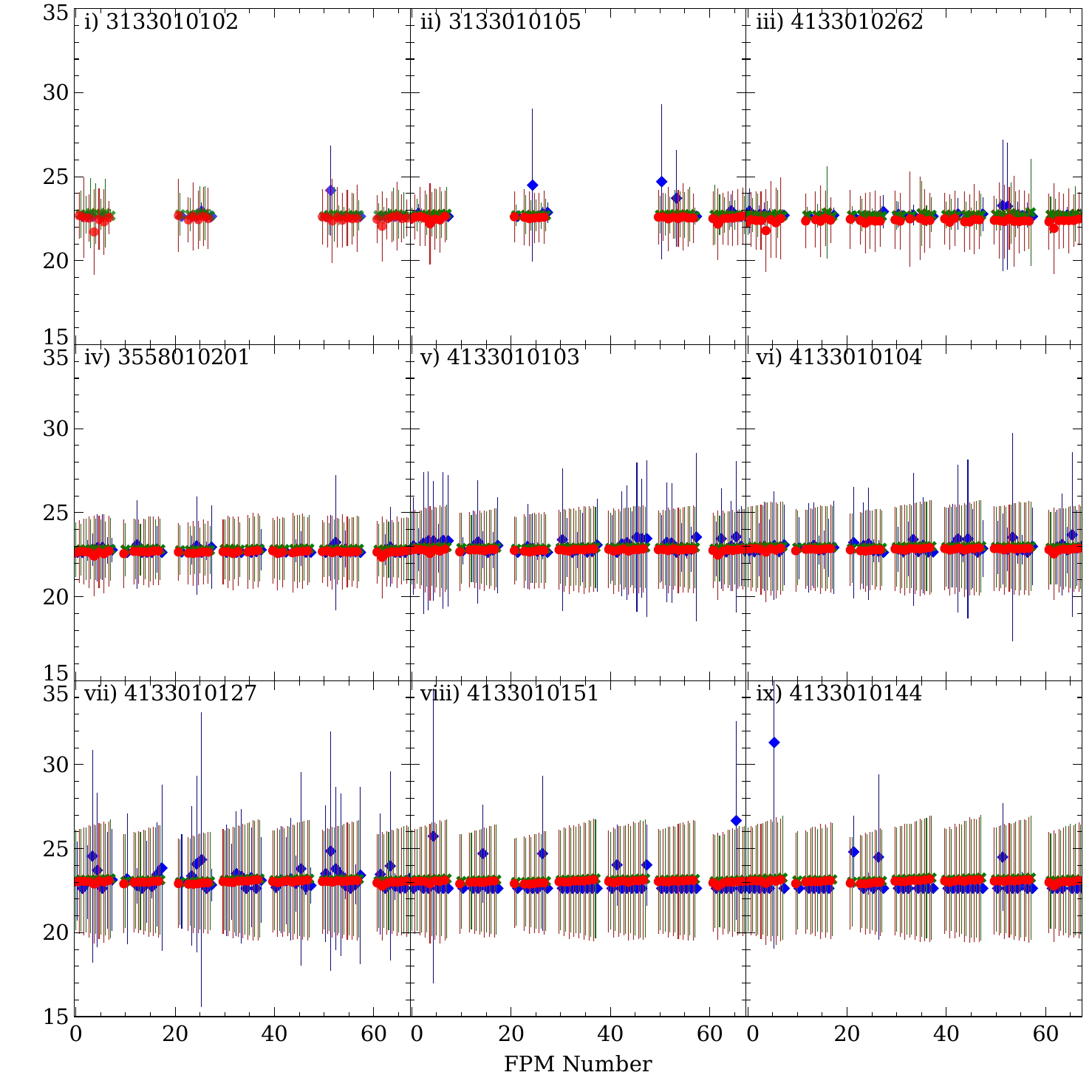}
\caption{Deadtime per event contributing to the lightcurve across the different FPMs in broad energy bands. Panels show the average deadtime per lightcurve event for the nine observations listed in Table \ref{tab:obs}. In each panel the low energy ($E<2.0\,\text{keV}$), medium energy ($2.0\,\text{keV}\leq E\leq10.0\,\text{keV}$), and high energy($E>10.0\,\text{keV}$) average deadtimes are displayed as red circles, green crosses and blue diamonds respectively. For some FPMs in all observations, except 4133010104, there were instances where the spread on deadtimes for individual FPMs in the highest energy band were 0. As such for these data points there are no error bars.}
\label{fig:dtperfpm-eband}
\end{figure*}

Finally, we consider the recorded deadtime per X-ray event across the detectors in different energy ranges. Fig. \ref{fig:dtperfpm-eband} shows the average deadtime per X-ray event across the active FPMs for all nine observations in the energy bands: 0.2--2.0keV, 2.0--10.0keV, 10.0--15.0keV. From this figure across the nine observations there does not appear to be any general trend in the deadtime per X-ray event in these different energy bands. There are isolated instances of FPMs, particularly in the highest energy band, being subject to unusually high average deadtime, such as FPM 5 in observation 4133010144, or unusually large spread, such as FPM 4 in observation 4133010151. In Table \ref{tab:deadtime-fpmebands} we display the count rates, minimum average deadtime per event per FPM, median average deadtime per event per FPM, and maximum average deadtime per event per FPM across the three energy bands and nine observations.

\begin{table*}
	\centering
	\caption{Median and most extreme values for the average deadtime associated with each X-ray event across FPMs for each of the nine observations listed. Values listed are the median of the average deadtimes per FPM associated with X-ray events in the energy bands stated.}
	\label{tab:deadtime-fpmebands}
	\begin{tabular}{l|cccc|cccc|cccr}
            \hline
            \multicolumn{1}{c}{} &
            \multicolumn{4}{c}{$E<2\text{keV}$ ($\mu s$)} &
            \multicolumn{4}{c}{$2\text{keV}\leq E\leq 10\text{keV}$ ($\mu s$)} &
            \multicolumn{4}{c}{$E>10\text{keV}$ ($\mu s$)} \\
		\hline
		OBSID & Rate (/s) & Min & Median & Max & Rate (/s) & Min & Median & Max & Rate (/s) & Min & Median & Max \\
		\hline
            3133010102 & 19.18 & 21.71 & $22.53$ & 22.69 & 12.11 & 22.65 & $22.72$ & 22.84 & 0.06 & 22.64 & $22.64$ & 24.19 \\
            3133010105 & 32.38 & 22.18 & $22.54$ & 22.65 & 19.77 & 22.66 & $22.73$ & 22.79 & 0.09 & 22.64 & $22.64$ & 24.70 \\
            4133010262 & 44.76 & 21.79 & $22.39$ & 22.53 & 26.63 & 22.63 & $22.69$ & 22.88 & 2.55 & 22.64 & $22.64$ & 23.28 \\
            3558010201 & 500.0 & 22.34 & $22.66$ & 22.75 & 248.7 & 22.76 & $22.82$ & 22.86 & 0.70 & 22.64 & $22.68$ & 23.23 \\
            4133010103 & 2418 & 22.47 & $22.78$ & 22.85 & 784.2 & 22.85 & $22.91$ & 22.95 & 0.78 & 22.64 & $22.92$ & 23.57 \\
            4133010104 & 2951 & 22.55 & $22.83$ & 22.92 & 983.9 & 22.88 & $22.95$ & 23.00 & 0.70 & 22.64 & $22.87$ & 23.68 \\
            4133010127 & 4906 & 22.77 & $23.04$ & 23.15 & 1874 & 23.00 & $23.13$ & 23.21 & 0.22 & 22.64 & $23.10$ & 24.85 \\
            4133010151 & 4984 & 22.76 & $23.04$ & 23.16 & 1957 & 23.01 & $23.14$ & 23.22 & 0.07 & 22.64 & $22.64$ & 26.67 \\
            4133010144 & 5029 & 22.78 & $23.06$ & 23.19 & 2000 & 23.02 & $23.17$ & 23.27 & 0.09 & 22.64 & $22.64$ & 31.32 \\
		\hline
	\end{tabular}
\end{table*}

In the lower and middle energy bands we do find that the minimum, median, and maximum averaged deadtime per X-ray event per FPM increases as the count rate increases. In the lowest energy band the median average deadtime per event increases by $2.99\%$ between the lowest and greatest values while the count rate increases by $11234\%$ in comparison for the energy band in those two observations. In the middle energy band the median average deadtime per X-ray event increases by $2.12\%$ between the lowest and greatest values while the count rate in that band for those two observations increases by $7509\%$. There is no such apparent trend in the greatest energy band, although that may be in part due to the significantly lower count rates in this band.

\subsection{Deadtime Effects on Timing Products}
\label{subsec:dteffects}

In order to examine the effects of deadtime on lightcurves of observations we consider the proportional deadtime in individual time bins in observations. We create lightcurves for all nine observations for the broad energy band 0.2--15.0keV, with a time bin size of 0.01s and we then determine the deadtime in each time bin as described in Section \ref{sec:methods}. This total deadtime in each bin is then corrected for the time bin size and the number of active detectors at that time to create values for the fractional deadtime in time bins. We then consider the fractional deadtime affecting individual time bins within these lightcurves. The average and most extreme values for this fractional deadtime are listed in Table \ref{tab:propdt}.
\begin{table*}
    \centering
    \caption{Proportional deadtime affecting lightcurves binned at a rate of 0.01s. The values quoted in the table are: the average count rate in the broad energy band for X-ray events between 0.2--15.0keV, minimum value for the proportional deadtime for any single bin in the lightcurve, mean and standard deviation on the mean value for the proportional deadtime for time bins in the lightcurve, and the maximum value for the proportional deadtime for any single bin in the lightcurve.}
    \label{tab:propdt}
    \begin{tabular}{lcccr}
    \hline
    Obs ID & Average Rate (/s) & Minimum & Average Fractional Deadtime & Maximum \\
    \hline
         3133010102 & 31.2 & 0.000000 & $0.000453\pm0.000290$ & 0.002523 \\
         3133010105 & 52.2 & 0.000000 & $0.000469\pm0.000281$ & 0.002083 \\
         4133010262 & 73.8 & 0.000000 & $0.000351\pm0.000265$ & 0.002106 \\
         3558010201 & 749 & 0.000076 & $0.000958\pm0.000324$ & 0.003131 \\
         4133010103 & 3202 & 0.000392 & $0.001668\pm0.000371$ & 0.003982 \\
         4133010104 & 3935 & 0.000581 & $0.001994\pm0.000399$ & 0.004295 \\
         4133010127 & 6780 & 0.001574 & $0.003332\pm0.000440$ & 0.005813 \\
         4133010151 & 6933 & 0.001320 & $0.003339\pm0.000439$ & 0.005482 \\
         4133010144 & 7024 & 0.001929 & $0.003594\pm0.000462$ & 0.006074 \\
         \hline
    \end{tabular}
\end{table*}

There is a clear trend for greater values for the fractional deadtime with increasing count rate. We can see this trend in the minimum, maximum and average fractional deadtime in time bins across the nine observations. This result is as expected, as a greater count rate should mean a greater amount of deadtime within each time bin, and as such a greater fractional deadtime. From these preliminary results it would also appear as though detector deadtime at count rates of the order of $\sim7000\,\text{cts}\slash s$ with 52 detectors is still limited below $0.01$, and it is possible that it won't have a serious effect even at the levels when telemetric deadtime becomes a factor.

\begin{figure*}
\centering
    \begin{subfigure}[]{0.3\textwidth}
         \centering
         \includegraphics[width=\textwidth]{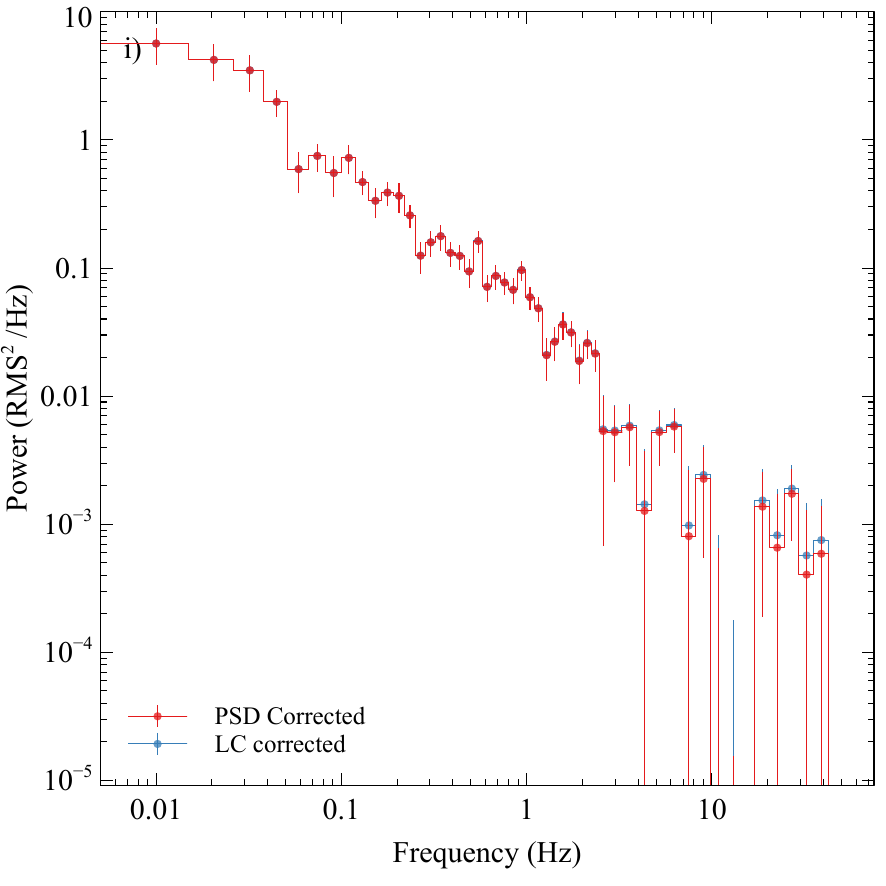}
         \caption{Observation 3133010102}
         \label{fig:psd1}
     \end{subfigure}
     \begin{subfigure}[]{0.3\textwidth}
         \centering
         \includegraphics[width=\textwidth]{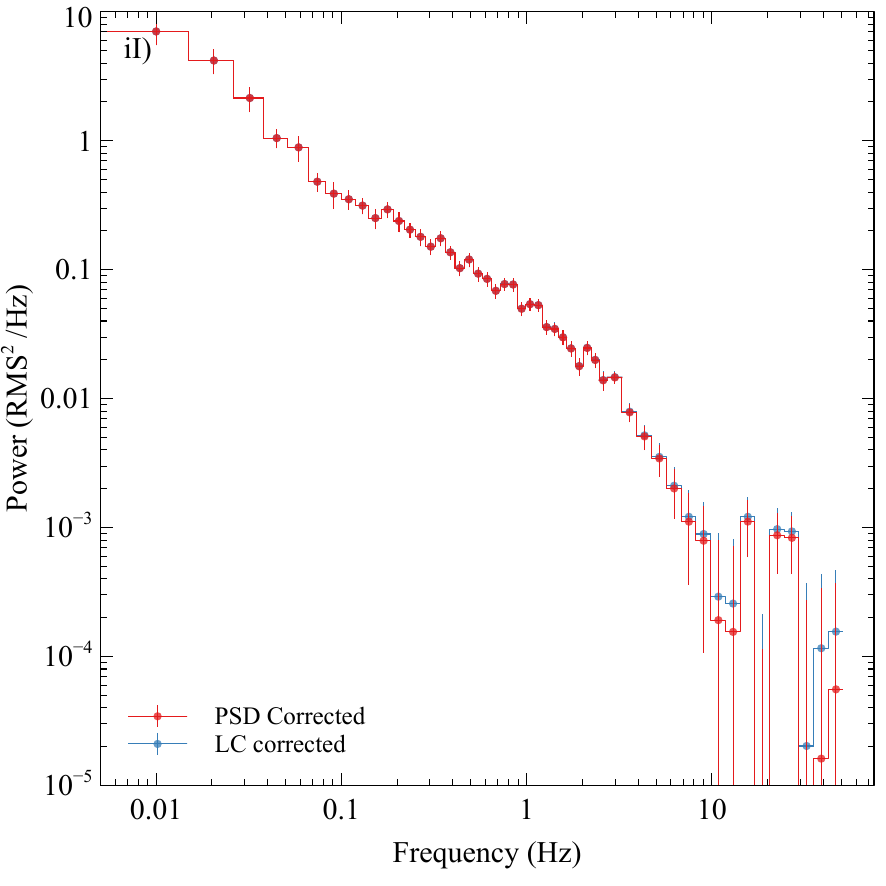}
         \caption{Observation 3133010105}
         \label{fig:psd2}
     \end{subfigure}
     \begin{subfigure}[]{0.3\textwidth}
         \centering
         \includegraphics[width=\textwidth]{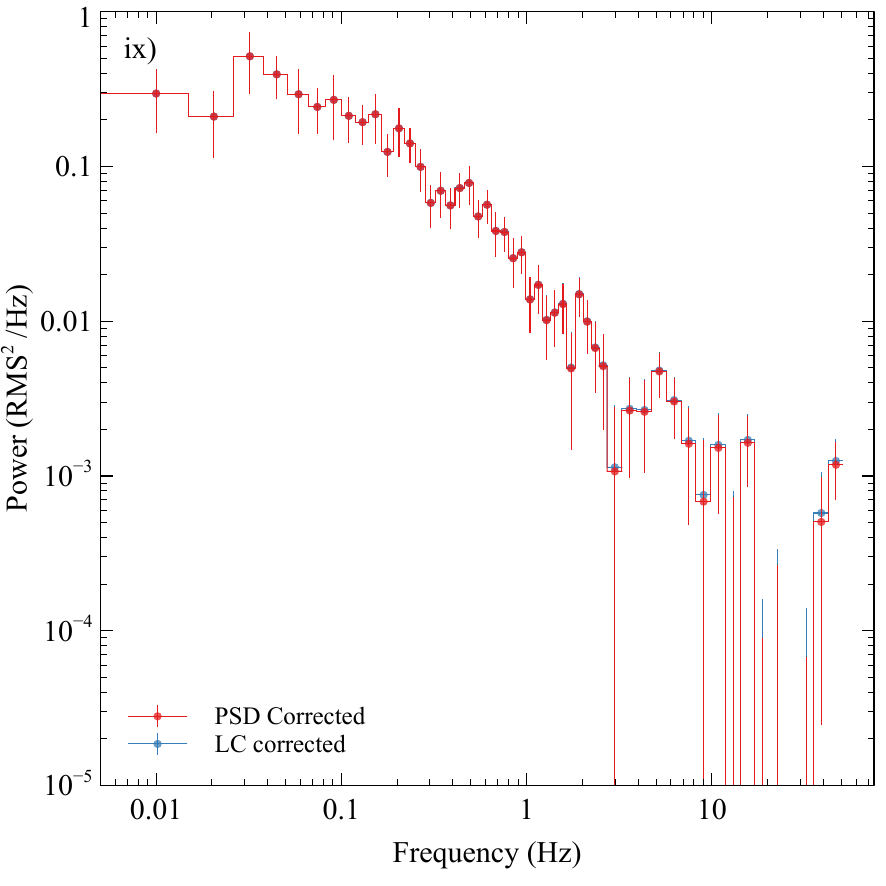}
         \caption{Observation 4133010262}
         \label{fig:psd9}
     \end{subfigure}
     \begin{subfigure}[]{0.3\textwidth}
         \centering
         \includegraphics[width=\textwidth]{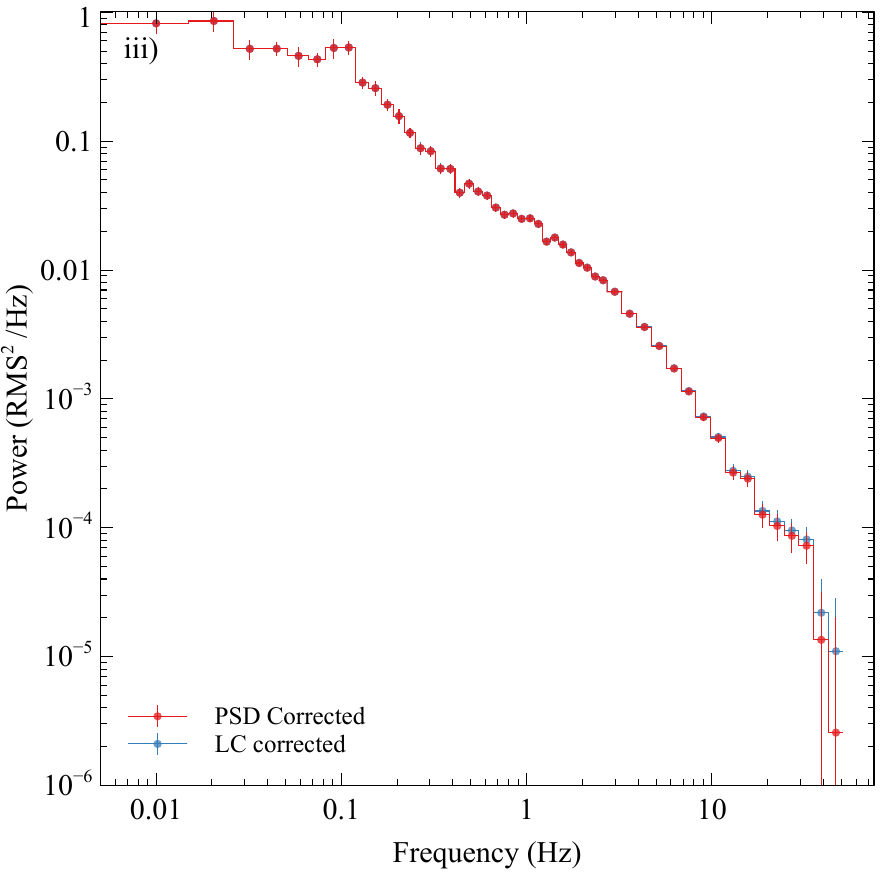}
         \caption{Observation 3558010201}
         \label{fig:psd3}
     \end{subfigure}
     \begin{subfigure}[]{0.3\textwidth}
         \centering
         \includegraphics[width=\textwidth]{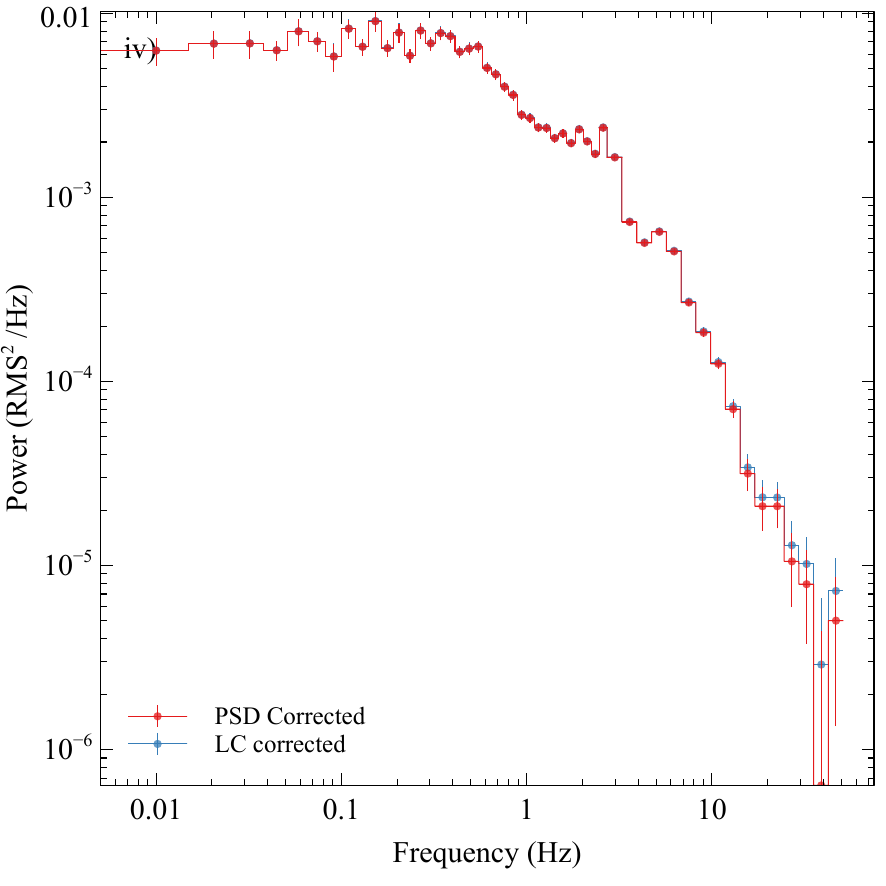}
         \caption{Observation 4133010103}
         \label{fig:psd4}
     \end{subfigure}
     \begin{subfigure}[]{0.3\textwidth}
         \centering
         \includegraphics[width=\textwidth]{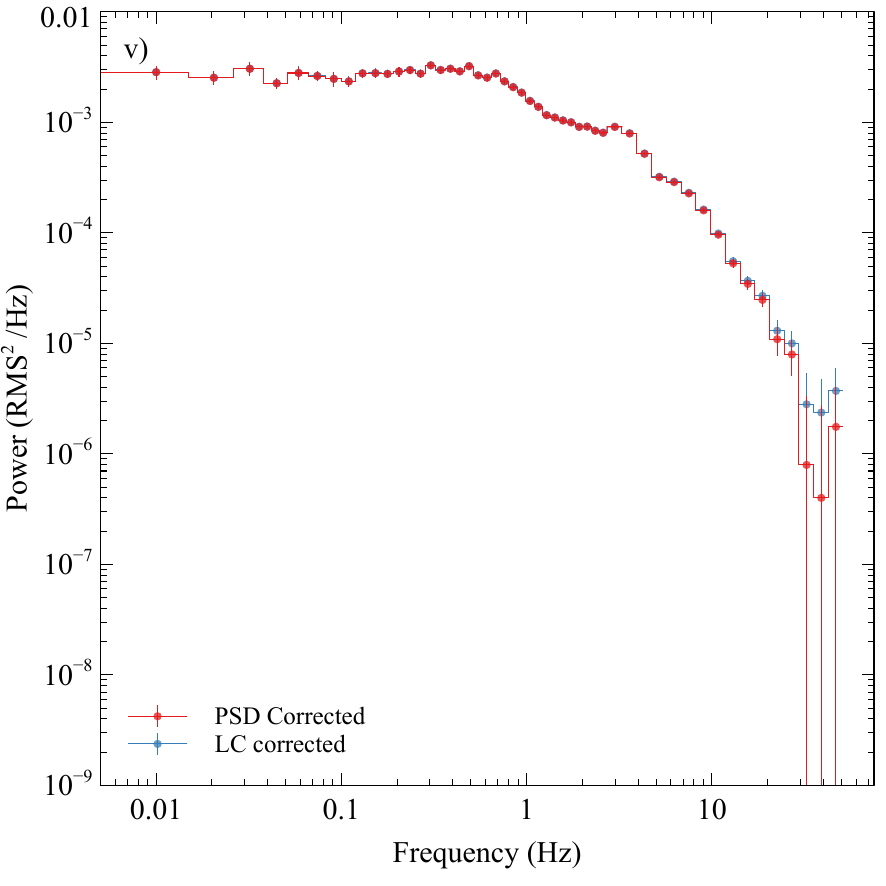}
         \caption{Observation 4133010104}
         \label{fig:psd5}
     \end{subfigure}
     \begin{subfigure}[]{0.3\textwidth}
         \centering
         \includegraphics[width=\textwidth]{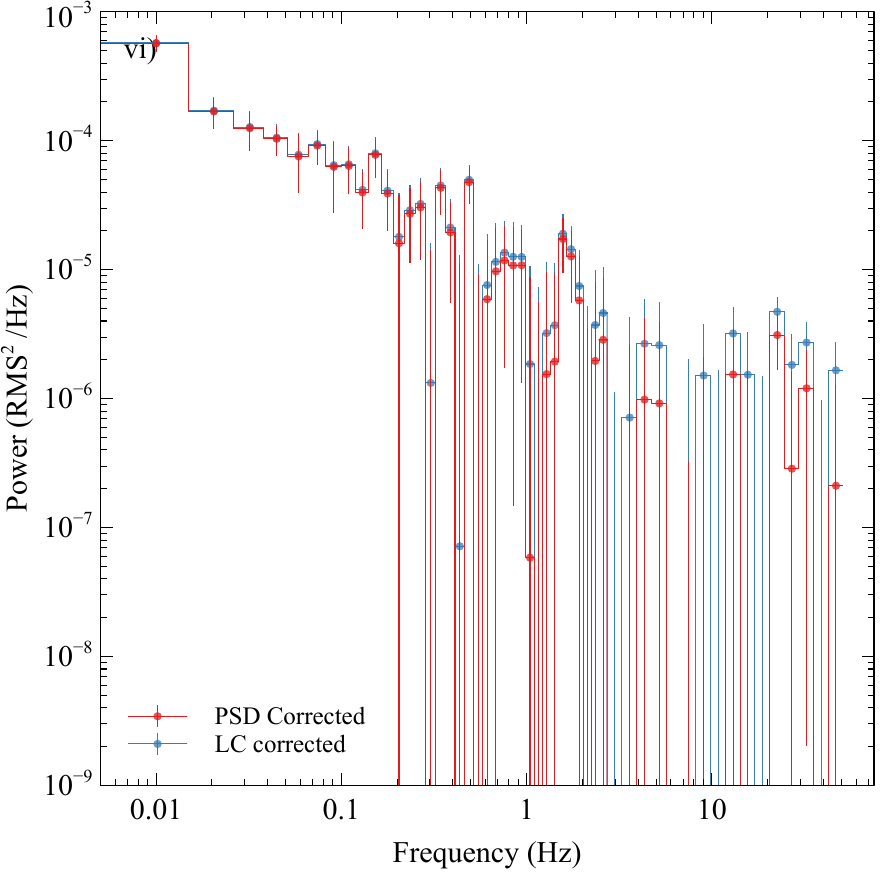}
         \caption{Observation 4133010127}
         \label{fig:psd6}
     \end{subfigure}
     \begin{subfigure}[]{0.3\textwidth}
         \centering
         \includegraphics[width=\textwidth]{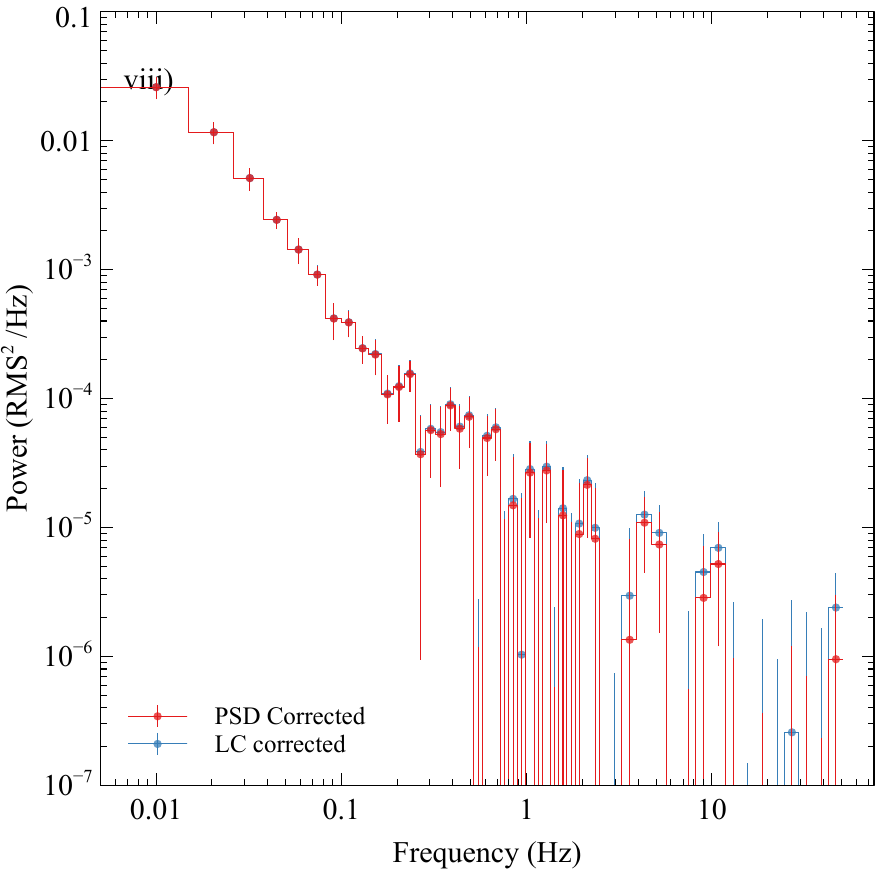}
         \caption{Observation 4133010151}
         \label{fig:psd8}
     \end{subfigure}
     \begin{subfigure}[]{0.3\textwidth}
         \centering
         \includegraphics[width=\textwidth]{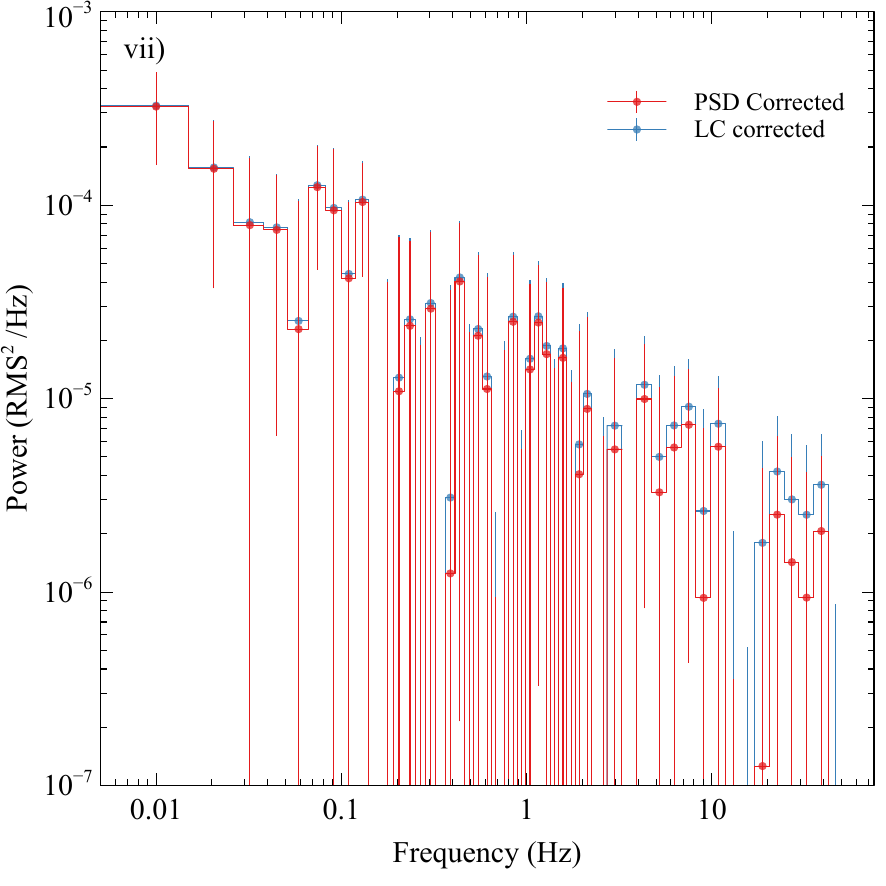}
         \caption{Observation 4133010144}
         \label{fig:psd7}
     \end{subfigure}
    \caption{Power spectra for observations of the X-ray binary GX 339-4. In all cases the power spectra have been created with lightcurves in the full energy range from 0.2--15.0\,keV, with 0.01\,s time binning and averaged over segments of length 100\,s. The red plot gives the power spectrum as corrected for noise as per Equation \ref{eq:zhang-noise}, and the blue plot gives the power spectrum as calculated with a flat Poisson noise level subtracted from a deadtime corrected lightcurve.}
    \label{fig:psds}
\end{figure*}

In Fig. \ref{fig:psds} we display the power spectra for these nine observations. These power spectra are recovered from the lightcurves created using photon energies in the range 0.2--15.0\,keV, with time binning of 0.01s and by averaging segments with length of 100s. We normalise these power spectra using the approach discussed by \cite{Miyamoto1992} and then subtract the Poisson noise. We create lightcurves using both the raw data and our deadtime corrected lightcurves, as discussed in Section \ref{sec:methods} in Equation \ref{eq:dtcorr}. From the power spectrum of the deadtime corrected lightcurve we subtract flat Poisson noise as $\frac{2}{r_{\text{e}}}$, and from the uncorrected lightcurve we subtract frequency-dependent noise as outlined in Equation \ref{eq:zhang-noise} \citep[][etc.]{nowak_rossi_1999}. In order to determine the very large event rate we considered the deadtimes of individual events detected during each observation. For the purposes of this investigation we consider very large events to be those which have a deadtime that is significantly greater than that of the average X-ray event. We set the deadtime threshold for consideration as a very large event at 3$\sigma$ from the mean deadtime value for X-ray events, and count the very large event rate accordingly. These power spectra are then rebinned in log space by a ratio of 0.1 up to 2.5Hz, and then by a ratio of 0.2 above 2.5Hz to account for noise at higher frequencies in some of the power spectra.
From an examination of the power spectra we can see that there are only minor differences at frequencies $f\lesssim1Hz$. Above this frequency there are differences between the two spectra, although these are usually within the error bars of the rebinned spectra. There does not appear to be any difference in the separation between the two spectra at the different count rates as displayed in these figures. We also note that in those spectra which are most affected by noise at the highest frequencies the spectrum from the deadtime corrected lightcurve still shows significantly noisy behaviour.

\section{Discussion}
\label{sec:discuss}
From Figures \ref{fig:dtperfpm} and \ref{fig:dtperfpm-eband} it appears that the deadtime per recorded event and per X-ray event is consistent across the different detectors. We fit the observed average deadtime per recorded event across detectors to a constant and find that for all observations the deadtime across detectors is consistent with a constant in all cases the reduced chi-square statistic showed that the averaged deadtime per event across FPMs was consistent with a constant value at more than 7$\sigma$. The average deadtime per event consistent with a minimised mean square error across FPMs, and the reduced chi square statistic for such values are given in Table \ref{tab:dtfpmfit}.

\begin{figure}
\centering
    \begin{subfigure}[]{0.45\textwidth}
         \centering
         \includegraphics[width=\textwidth]{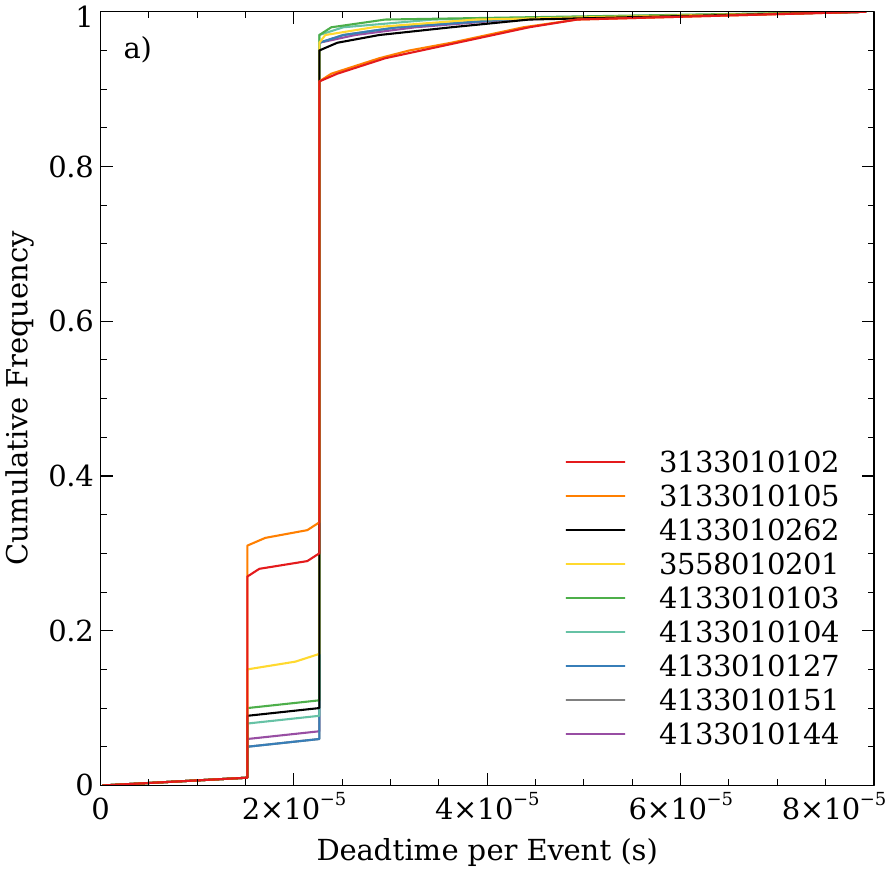}
         \caption{All Events}
         \label{fig:dtcfuf}
     \end{subfigure}
     \begin{subfigure}[]{0.45\textwidth}
         \centering
         \includegraphics[width=\textwidth]{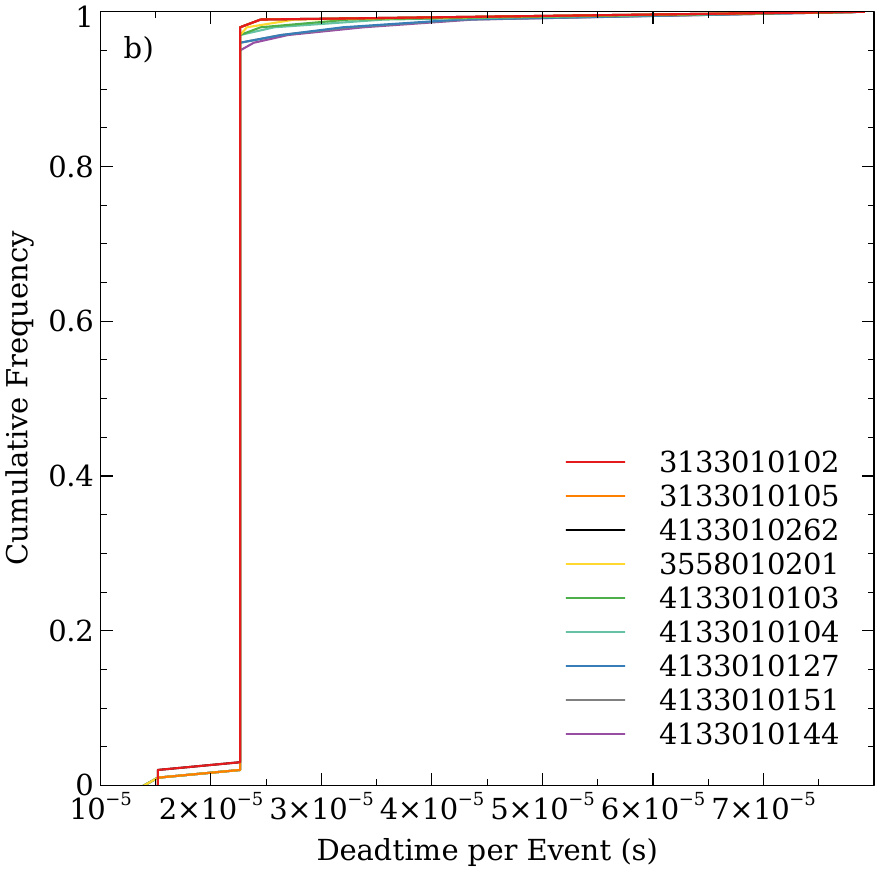}
         \caption{X-ray Events}
         \label{fig:dtcfcl}
     \end{subfigure}
    \caption{Distribution of deadtimes for individual events for the nine observations analysed. Deadtimes are as recorded in a) the unfiltered and unscreened event file, and b) the cleaned and screened X-ray event file.}
    \label{fig:evtdt}
\end{figure}

\begin{figure}
\includegraphics[width=0.9\columnwidth]{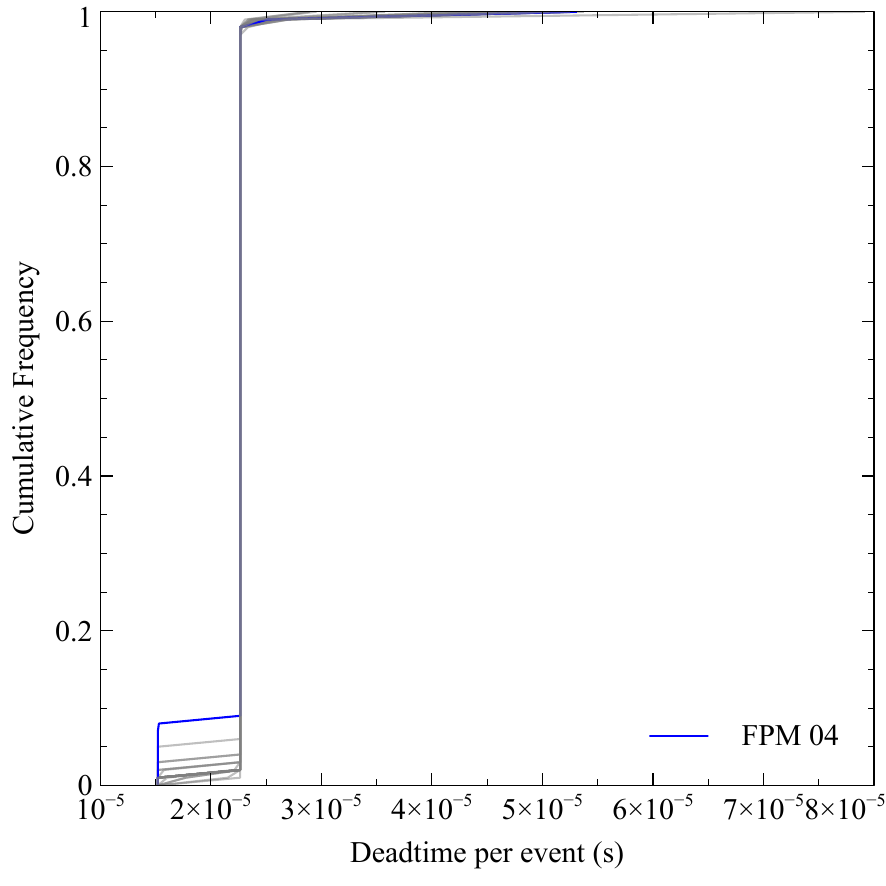}
\caption{Distributions of X-ray event deadtimes for the different FPMs in observation 3133010102 of GX 339-4. Events have energies in the range 0.2--15.0keV. The labelled blue line shows the one FPM, 04, which has a statistically significant different distribution of event deadtimes when compared with the other 27 FPMs which were active during the Good Time Intervals for this observation (in grey) which were all consistent with each other.}
\label{fig:o1perfpmdt}
\end{figure}

\begin{table}
    \centering
    \caption{Constant values for deadtime per event across FPMs, and goodness of fit statistic. For all observations there are 51 degrees of freedom.}
    \label{tab:dtfpmfit}
    \begin{tabular}{lcr}
    \hline
    Obs ID & Deadtime per event (/s) & Red. Chi Square \\
    \hline
         3133010102 & $2.2626 \times 10^{-5}$ & 0.0841 \\
         3133010105 & $2.2733 \times 10^{-5}$ & 0.0595 \\
         4133010262 & $2.2444 \times 10^{-5}$ & 0.0102 \\
         3558010201 & $2.1844 \times 10^{-5}$ & 0.0072 \\
         4133010103 & $2.2079 \times 10^{-5}$ & 0.0045 \\
         4133010104 & $2.2298 \times 10^{-5}$ & 0.0028 \\
         4133010127 & $2.2686 \times 10^{-5}$ & 0.0013 \\
         4133010151 & $2.2694 \times 10^{-5}$ & 0.0012 \\
         4133010144 & $2.2637 \times 10^{-5}$ & 0.0010 \\
         \hline
    \end{tabular}
\end{table}
In all observations the reduced chi squared value suggests that the constant value across detectors within single observations provides an adequate model. This would imply that no single detector is subject to a substantially higher proportion of large deadtime events. This is in agreement with the work of \cite{vivekanand_phase-resolved_2021} when a similar analysis was performed regarding observations of the Crab pulsar, where they found that, with the exception of one detector (00), the detector deadtime across FPMs was consistent with a constant value. The very low reduced chi-square totals indicate that the model is overfitting to the data, even as a simple constant value. This is due to the vast majority of events being recorded at specific discrete values for the dead time, particularly for events which are considered to be `good' X-ray events.
When we consider the distribution of deadtimes recorded for all events across the whole duration of an observation, we find that each observation has a distinct distribution of different deadtimes recorded per event. We perform Kolmogorov-Smirnov tests for every pair of observations and find that the null hypothesis that each pair of distributions is drawn from the same distribution can be rejected at the 0.1$\%$ level for each pair of observations independently. However when we consider the deadtimes only for those events which pass all of the filtering and screening, there are several pairs of observations which have similar event deadtime distributions. We fail to reject at the 10$\%$ level that observations 3133010102 and 3133010105, 3133010102 and 3558010201, 3133010102 and 4133010262, and 4133010127 and 4133010151 have the same underlying deadtime per X-ray event distributions. In Fig. \ref{fig:evtdt} we illustrate the cumulative frequency curves for the deadtimes of all recorded events and for X-ray events for these nine observations. In all cases the deadtimes for events are split into three broad bands, being less than $\sim$24$\mu$s, from $\sim$24--27$\mu$s, and greater than $\sim$27$\mu$s. There are small populations of high deadtime events, and the proportions of small deadtime events range between $\sim$5 and 10\% of the total populations. Additionally, the distribution of deadtimes per event on a per-FPM basis within individual observations are statistically distinct when conducting pair-wise Kolmogorov-Smirnov tests with any pair of FPMs and can reject the null hypothesis for any pair at the 0.1$\%$ level that the distributions are the same. On a per-FPM basis within observations this trend becomes less distinct when looking again at the cleaned and screened X-ray event lists. In all cases there are combinations of FPMs where we cannot reject even at the 10$\%$ level that the deadtime distributions per event per FPM are being drawn from the same distribution. For observation 3133010102 all of the FPMs with the exception of FPM 04 have statistically indistinct deadtime per event distributions, and we illustrate this in Fig. \ref{fig:o1perfpmdt}. Pair-wise Kolmogorov-Smirnov testing showed that FPM 04 was significantly different to all other FPMs (except FPM 62) at least at a 10\% confidence level, and in the case of 18 of the FPMs at less than the 1\% level, for this particular observation, a result not seen for any other FPM in any other observation.


These results indicate that even thought the ensemble effects of deadtime on detectors within observations is averaged out over long observing periods there can still be substantial differences in how the individual detectors are experiencing that deadtime. As such it is important to determine what effects, if any, there are on timing and Fourier-domain products.

\begin{figure*}
\includegraphics[width=\textwidth]{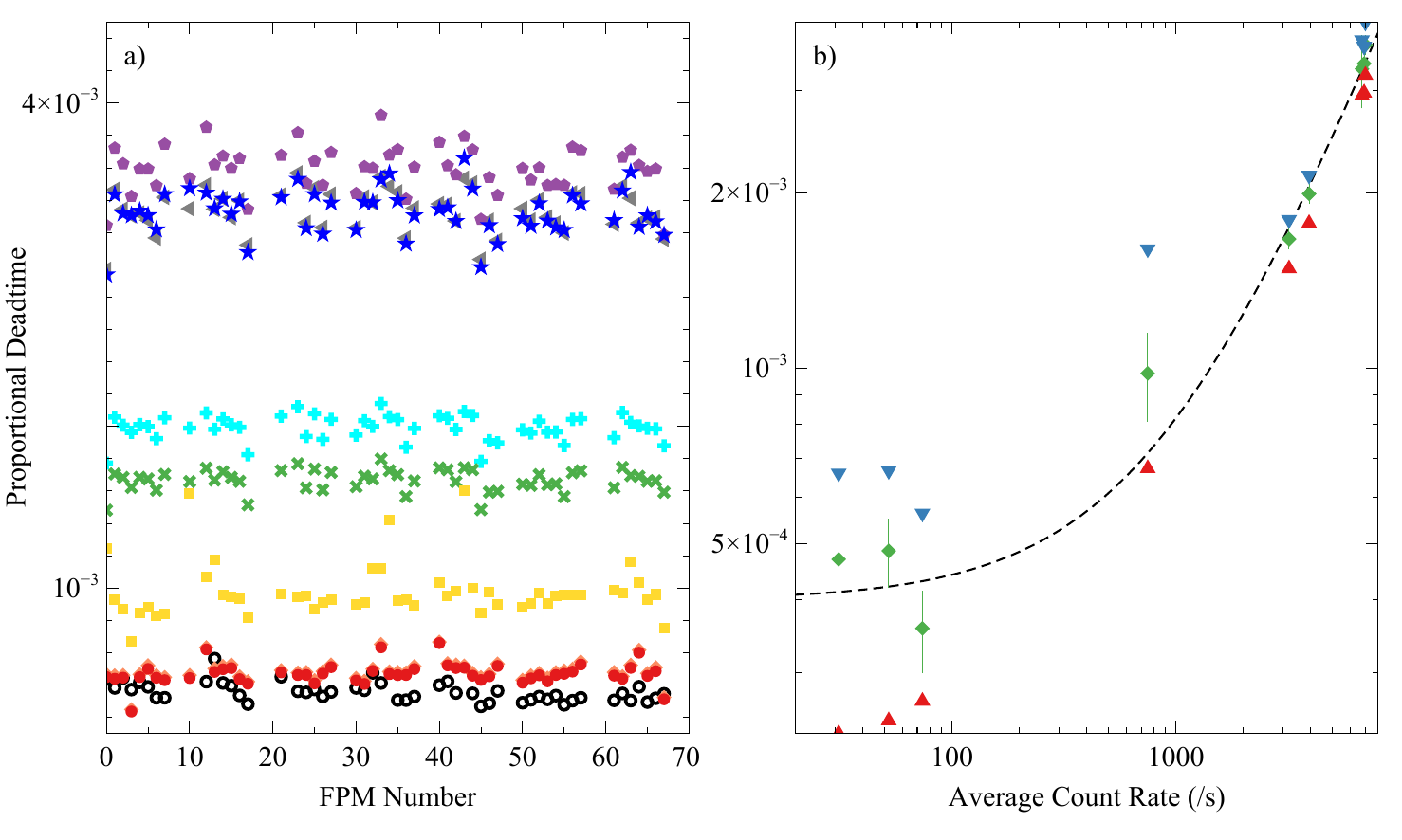}
\caption{Proportional deadtime experienced by individual FPMs during the nine observations being considered. In a) we display the proportional deadtime experienced by each FPM (3133010102 - red circle, 3133010105 - orange diamond, 3558010201 - yellow square, 4133010103 - green cross, 4133010104 - cyan plus, 4133010127 - blue star, 4133010144 - purple pentagon, 4133010151 - grey triangle, 4133010262 - black ring). In b) we show the minimum, average and standard deviation, and maximum value for the proportional deadtime experienced by the FPMs in the nine observations against the average count rate for each observation. The black dashed line shows the best fitting linear relationship.}
\label{fig:lcfracdt}
\end{figure*}

We first consider the effects of proportional deadtime at different recorded count rates on the lightcurves observed for these nine observations across the FPMs. The proportional deadtime experienced by the individual FPMs in each observation is variable, but significantly less variable than that experienced by the same FPM when the source count rate is changing. In panel (a) of Fig. \ref{fig:lcfracdt} we show the proportional deadtime experienced over the course of an observation by each FPM. We note that for all for all detectors in all observations the proportional deadtime is less than 0.5$\%$, even for our higher count rate observations. In assuming that there should be a constant proportional deadtime across all detectors we find that in the highest count rate observations (those where $r_{\text{Obs}} > 1000$ct/s) all of the proportional deadtimes for the individual FPMs are within $3\sigma$ of the mean. For those observations with lower count rates there was one FPM in each case with a proportional deadtime during the GTIs which experienced particularly high or low proportional deadtime. During observations 3133010102 and 3133010105 FPM 03 experienced significantly low proportional deadtimes, at 3.48 and 3.54$\sigma$ below the mean, while in observations 3558010201 and 4133010262 FPMs 10 and 13 experienced significantly higher proportional deadtimes at 4.62 and 3.51$\sigma$ above the mean respectively. In the cases of FPMs 10 and 13 in these two observations the total counts detected by these FPMs, including outside of the GTIs is significantly greater than that experienced by other detectors, on the order of $\sim 3 - 4$ times greater. As we might expect there is an increase in the proportional deadtime experienced by FPMs with increasing count rate, and the relationship between count rate and proportional deadtime is displayed on panel (b) of Fig. \ref{fig:lcfracdt}.
For a simple linear relationship with 7 degrees of freedom we find that such a relationship of the form
\begin{equation}
\tau_{\text{Dead,prop}} = a + b \times r_{\text{Obs}}
\label{eq:propdtvsrate}
\end{equation}
with $a=0.000399$ and $b=4.20\times10^{-7}$s, gives a reasonable quality of fit, with $\chi^2_{\nu} = 1.4772$ and seven degrees of freedom. At this quality of fit we are only able to reject the model at the 20\% level. Such a relationship would imply that detector deadtime as a proportion of time bins would require an observed count rate of $\sim23000$ before the cumulative effect is enough to cause a proportional deadtime of $1\%$. This may not be the case, however, as increasing count rate would increase the proportional deadtime, and as such the observed count rate would be suppressed. In this regime we would also expect to see the effects of telemetric deadtime, as has already been noted, and as such it is possible that detector deadtime will never have a significant effect on an observation which is also free of the effects of telemetric deadtime.

\begin{figure*}
\centering
    \begin{subfigure}[]{0.3\textwidth}
         \centering
         \includegraphics[width=\textwidth]{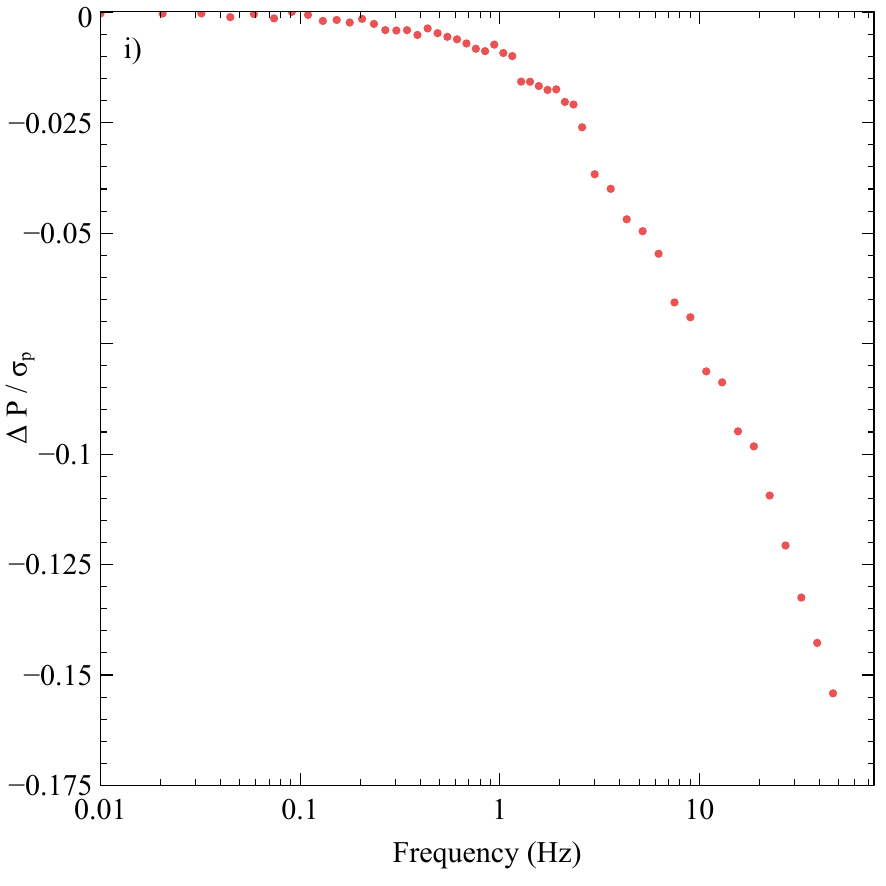}
         \caption{Observation 3133010102}
         \label{fig:psdd1}
     \end{subfigure}
     \begin{subfigure}[]{0.3\textwidth}
         \centering
         \includegraphics[width=\textwidth]{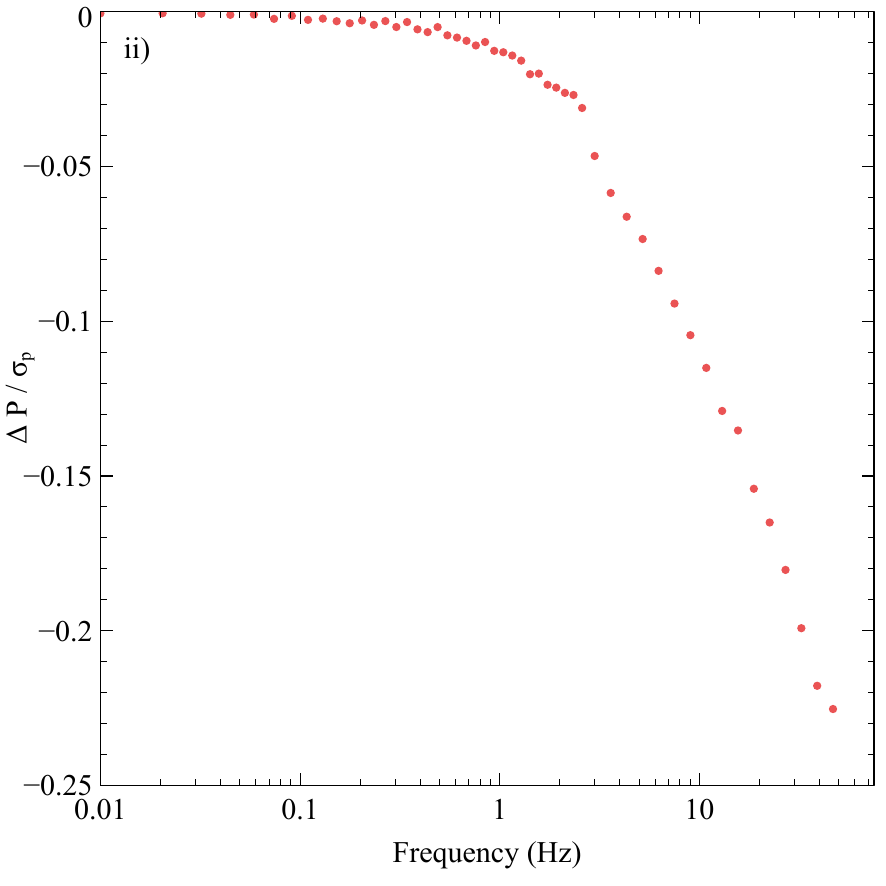}
         \caption{Observation 3133010105}
         \label{fig:psdd2}
     \end{subfigure}
     \begin{subfigure}[]{0.3\textwidth}
         \centering
         \includegraphics[width=\textwidth]{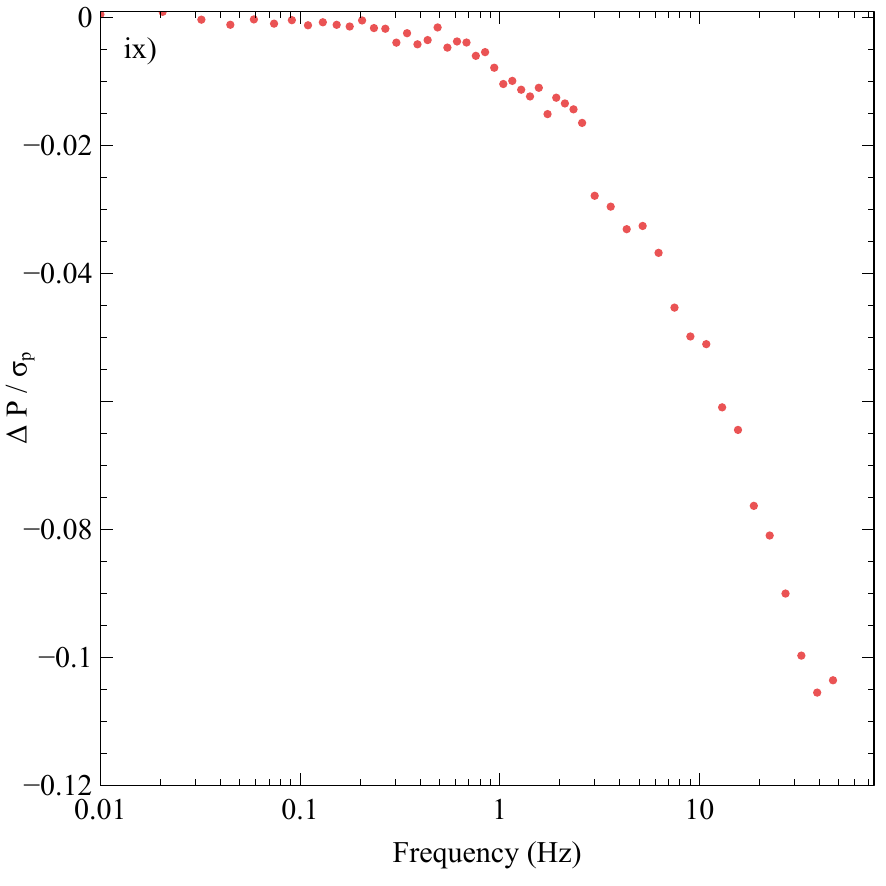}
         \caption{Observation 4133010262}
         \label{fig:psdd9}
     \end{subfigure}
     \begin{subfigure}[]{0.3\textwidth}
         \centering
         \includegraphics[width=\textwidth]{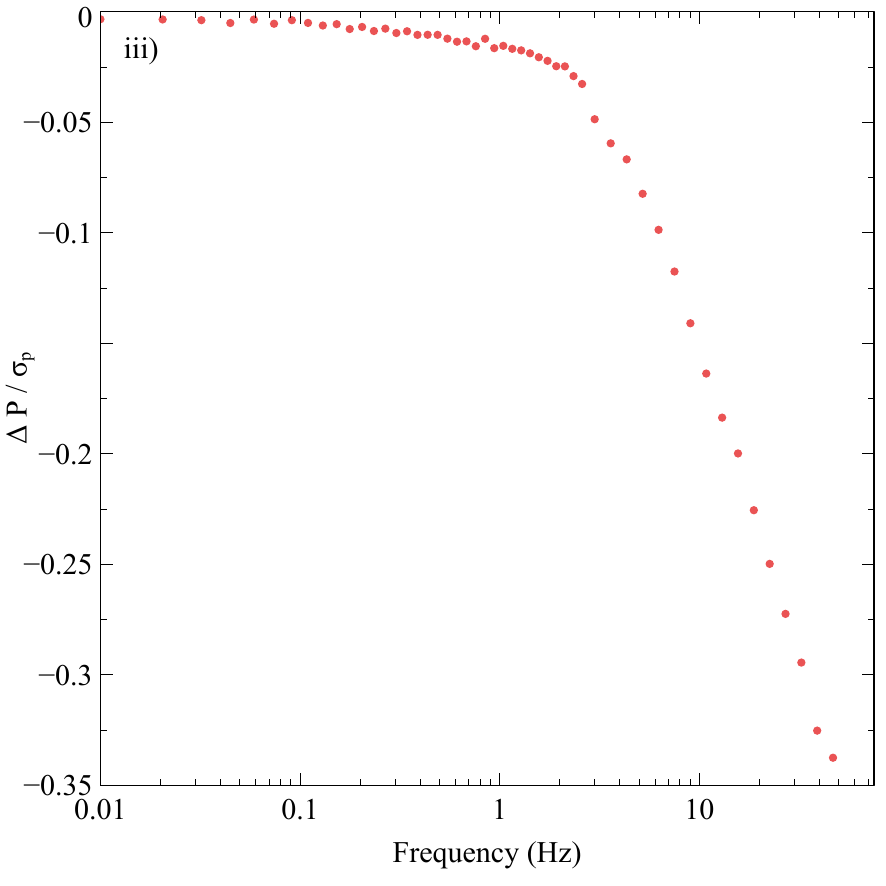}
         \caption{Observation 3558010201}
         \label{fig:psdd3}
     \end{subfigure}
     \begin{subfigure}[]{0.3\textwidth}
         \centering
         \includegraphics[width=\textwidth]{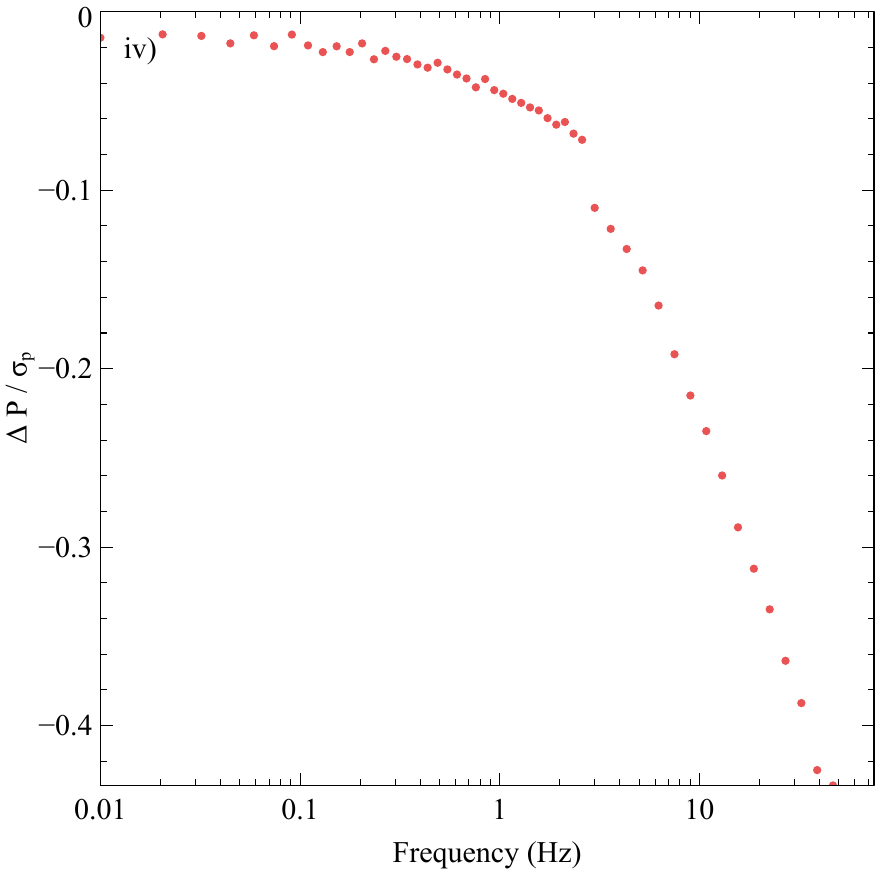}
         \caption{Observation 4133010103}
         \label{fig:psdd4}
     \end{subfigure}
     \begin{subfigure}[]{0.3\textwidth}
         \centering
         \includegraphics[width=\textwidth]{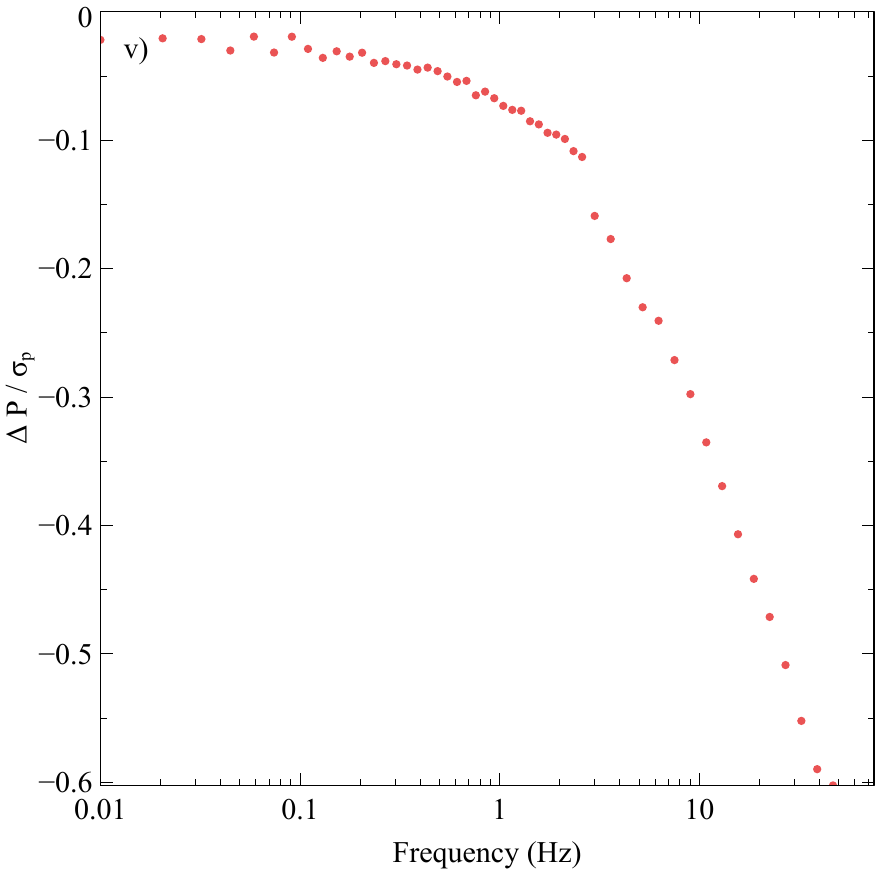}
         \caption{Observation 4133010104}
         \label{fig:psdd5}
     \end{subfigure}
     \begin{subfigure}[]{0.3\textwidth}
         \centering
         \includegraphics[width=\textwidth]{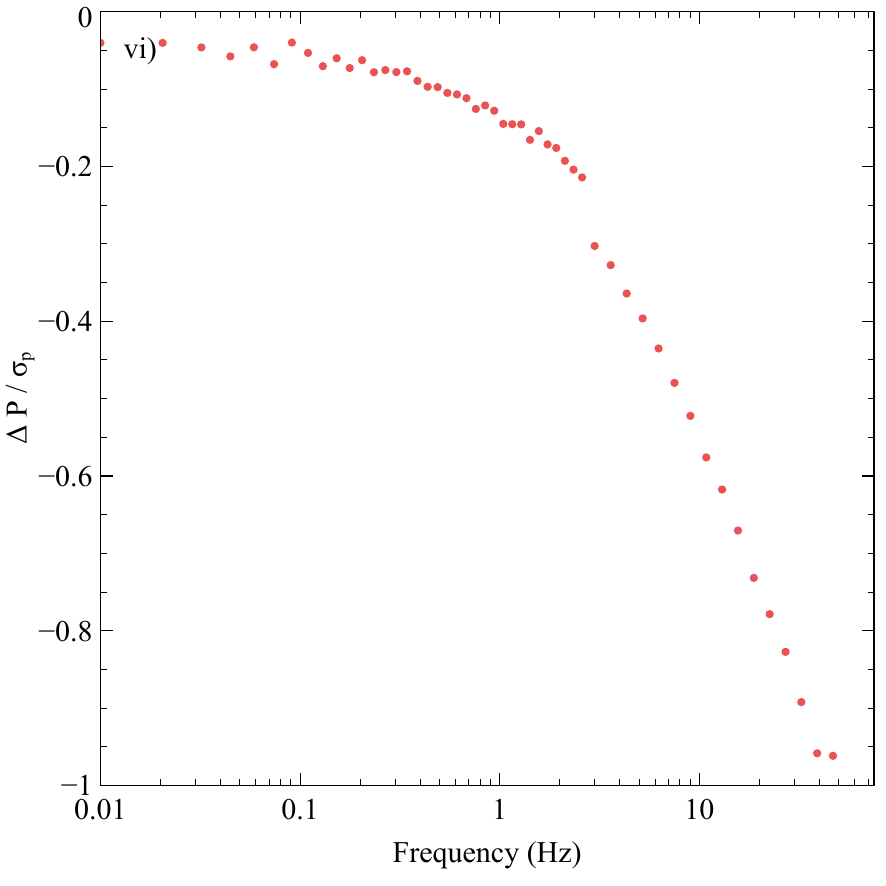}
         \caption{Observation 4133010127}
         \label{fig:psdd6}
     \end{subfigure}
     \begin{subfigure}[]{0.3\textwidth}
         \centering
         \includegraphics[width=\textwidth]{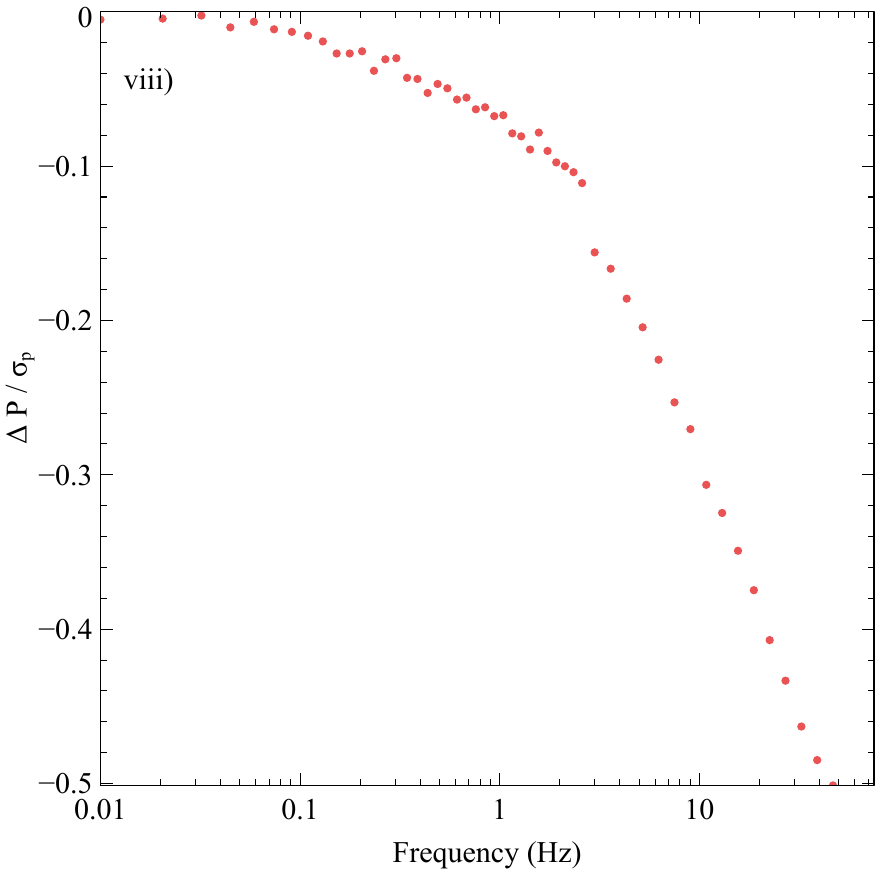}
         \caption{Observation 4133010151}
         \label{fig:psdd8}
     \end{subfigure}
     \begin{subfigure}[]{0.3\textwidth}
         \centering
         \includegraphics[width=\textwidth]{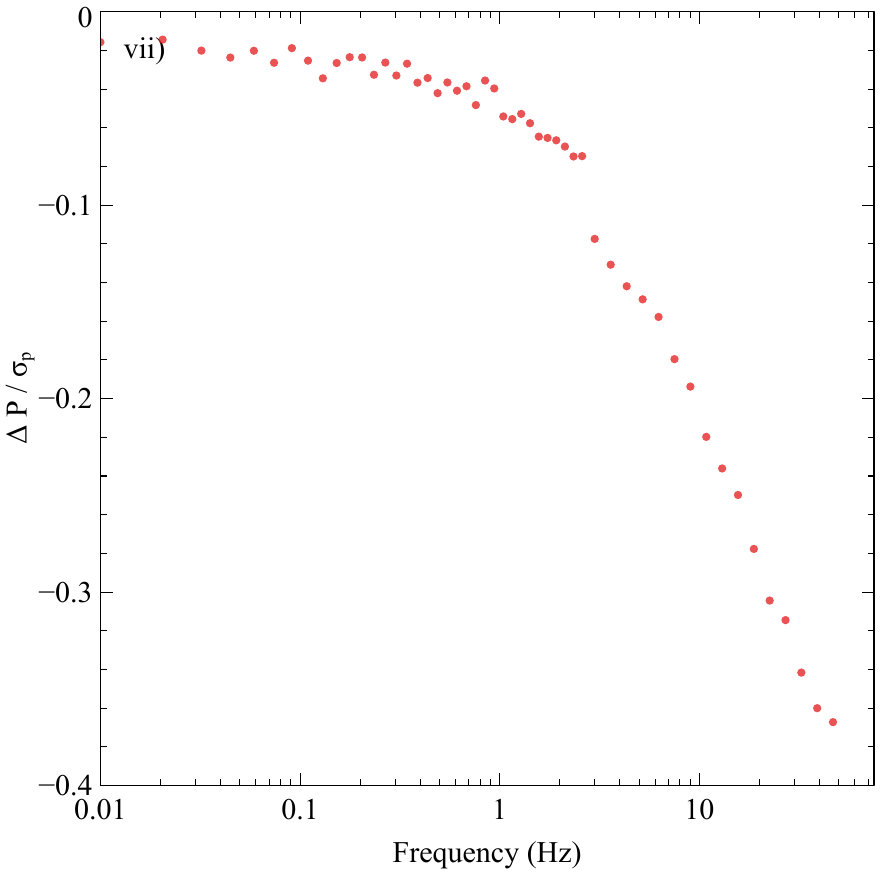}
         \caption{Observation 4133010144}
         \label{fig:psdd7}
     \end{subfigure}
     \caption{Deviations in the power spectrum for observations of the X-ray Binary GX 339-4 as calculated by a deadtime correction to the power spectrum as per Equation \ref{eq:zhang-noise}, and by making a deadtime correction to the lightcurve. The vertical axis is scaled according to the deviations in power between the two prescriptions compared with $\sigma_p$ at those frequencies. These are equivalent to the deviations in the red and blue plots in Fig. \ref{fig:psds}.}
    \label{fig:psddev}
\end{figure*}

Now that the scale of the effects on lightcurves has been determined we can examine the power spectra to see how this is propagated through to Fourier products. In Fig. \ref{fig:psds} we can see the averaged power spectra for the nine observations, and in all cases the two spectra are clearly very closely matched to frequencies of the order $\sim1$Hz. When we consider the deviations between the two models a very clear pattern emerges, and this is displayed in Fig. \ref{fig:psddev}.

As the frequency increases there is a clear trend towards an increasing deviation between the two power spectra. In all cases this deviation is negative, with the power from the deadtime corrected lightcurve being greater than the power from the power spectrum with frequency-dependent noise subtracted as in equation \ref{eq:zhang-noise}. For the lower count rate observations the magnitude of the deviation is smaller, and in the greatest case, in observation 4133010127, the deviation approaches the size of the error in the measurement. As such the differences in the two approaches do not appear to be statistically significant in these cases.

\begin{figure}
\includegraphics[width=0.9\columnwidth]{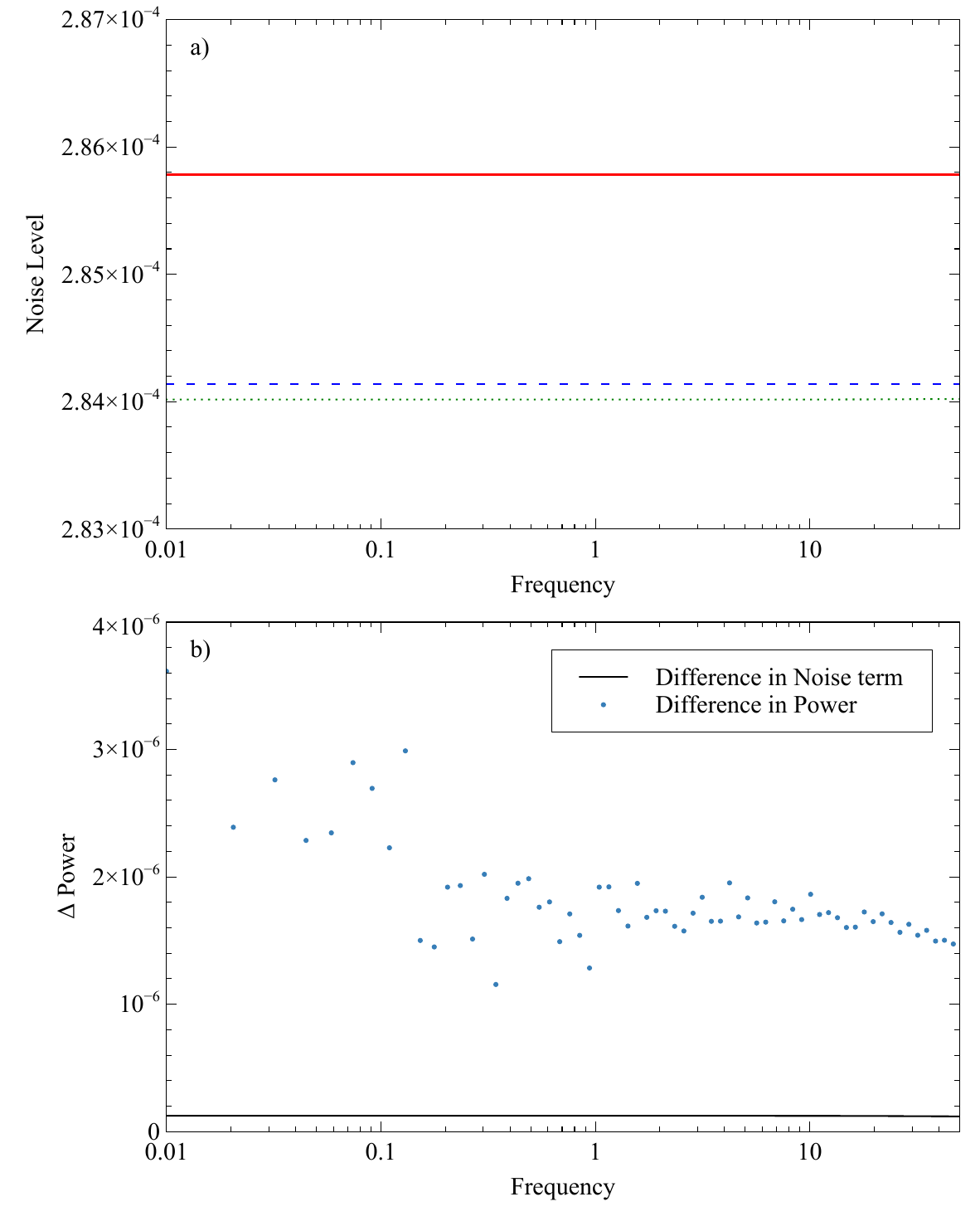}
\caption{Poisson noise subtraction to power spectra of observation 4133010144 of GX 339-4. The noise levels are calculated as: frequency-independent classical (2/r) noise (red solid line); frequency-independent classical noise calculated from a deadtime corrected lightcurve (blue dashed line); frequency-dependent Poisson noise as defined in Eq. \ref{eq:zhang-noise} (green dotted line). The lower panel shows the difference in the frequency-independent classical noise calculated from a deadtime corrected lightcurve and the frequency-dependent Poisson noise as defined in Eq. \ref{eq:zhang-noise} (black line), and the difference in power for the spectra corrected in each case (blue circles). In all cases power is to be subtracted from an averaged power spectrum created from segments of 100s of a lightcurve binned at 0.01s.}
\label{fig:noise-comp1}  
\end{figure}

\begin{figure}
\includegraphics[width=0.9\columnwidth]{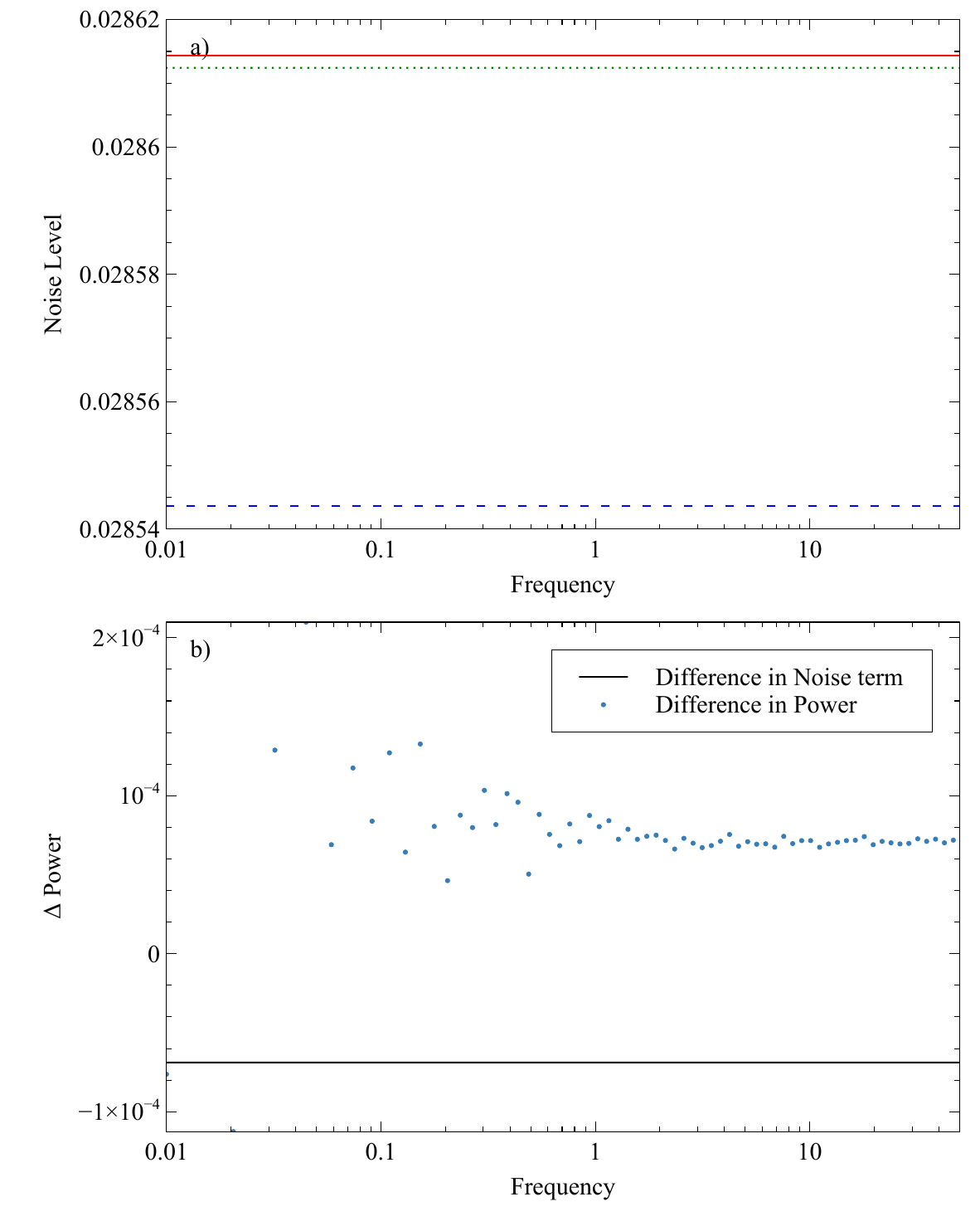}
\caption{Poisson noise subtraction to power spectra of observation 4133010262 of GX 339-4. The noise levels are calculated as: frequency-independent classical (2/r) noise (red solid line); frequency-independent classical noise calculated from a deadtime corrected lightcurve (blue dashed line); frequency-dependent Poisson noise as defined in Eq. \ref{eq:zhang-noise} (green dotted line). The lower panel shows the difference in the frequency-independent classical noise calculated from a deadtime corrected lightcurve and the frequency-dependent Poisson noise as defined in Eq. \ref{eq:zhang-noise} (black line), and the difference in power for the spectra corrected in each case (blue circles). In all cases power is to be subtracted from an averaged power spectrum created from segments of 100s of a lightcurve binned at 0.01s.}
\label{fig:noise-comp2}  
\end{figure}

In considering the apparent deviations in the two power spectra we now look at the frequency-dependent terms in Eq. \ref{eq:zhang-noise}. Across the nine observations the $t_{\text{b}}$ and $N_f$ terms will be constant (as we use a common time binning and segment length for averaged power spectra), but the other variables can vary. We distinguish between $\tau_{\text{d}}$ and $\tau_{\text{vle}}$ using the deadtimes in the unfiltered and screened event lists. We first determine $\tau_{\text{d}}$ by finding the average X-ray event deadtime from the screened and filtered event list. For the purposes of this analysis we then define very large events, and subsequently $\tau_{\text{vle}}$, as any event which has a deadtime which is more than $\tau_{\text{d}} + 3\times \sigma_{\tau_{\text{d}}}$. Very large events are then pulled from the unfiltered and un-screened raw event lists. In Table \ref{tab:zhang-vals} we list the values for those variables which can vary between observations.
\begin{table}
    \centering
    \caption{Values of variables to be passed to Eq. \ref{eq:zhang-noise} in calculating the frequency-dependent noise to be subtracted from power spectra.}
    \label{tab:zhang-vals}
    \begin{tabular}{lccccr}
    \hline
    Obs ID & $r_{\text{e}}$ (/s) & $r_{\text{pe}}$ (/s) & $\tau_{\text{d}}$ ($\mu$s) & $r_{\text{vle}}$ (/s) & $\tau_{\text{vle}}$ ($\mu$s) \\
    \hline
         3133010102 & 31.2 & 1.11 & 22.66 & 5.40 & 34.11 \\
         3133010105 & 52.2 & 1.79 & 22.66 & 5.81 & 34.33 \\
         4133010262 & 73.8 & 1.57 & 22.61 & 3.44 & 34.84 \\
         3558010201 & 749 & 15.0 & 22.74 & 17.7 & 38.48 \\
         4133010103 & 3202 & 61.6 & 22.84 & 40.6 & 41.49 \\
         4133010104 & 3935 & 75.7 & 22.89 & 56.6 & 41.91 \\
         4133010127 & 6780 & 130 & 23.08 & 140 & 43.50 \\
         4133010151 & 6933 & 133 & 23.08 & 141 & 43.52 \\
         4133010144 & 7024 & 135 & 23.11 & 163 & 43.48 \\
         \hline 
    \end{tabular}
\end{table}
When we examine the three terms in the frequency-dependent Poisson noise correction we find that for all observations the deviation from $2/r_{\text{e}}$ noise is greatest in the first term, and is at least of the order $10^2$ greater than the cosine and sine dependent terms in all cases. We expect these terms to be lower for \emph{NICER} than for previous instruments due to the large number of detectors onboard. \emph{NICER} regularly has up to 50 detectors functional, whereas \emph{RXTE} only had 5. In addition, while the deadtimes for X-ray events are comparable \citep[of the order $20\mu s$ for \emph{RXTE};][]{nowak_rossi_1999}, we find that the deadtime for large events is also lower for \emph{NICER} \citep[of the order $\sim40\mu s$ as opposed to $\sim70\mu s$;][]{nowak_rossi_1999}. Comparatively the $r_{\text{pe}}$ terms will be 10 times smaller for \emph{NICER} than \emph{RXTE}, and $\tau_{\text{vle}}$ will be smaller by a factor of nearly two. The relative decrease in $r_{\text{pe}}$ for similar count rates will mean that for identical observations the frequency-dependent noise will be far closer to the ``ideal'' $2\slash r_{\text{e}}$ Poisson noise. In Fig. \ref{fig:noise-comp1} we show a comparison of the Poisson noise levels subtracted from power spectra for observation 4133010144 as the highest count rate observation, where the effects of deadtime should be most evident.
In the high count rate regime we find that the noise calculated from a deadtime corrected lightcurve mirrors the frequency-dependent noise far closely than Poisson noise without any such correction, as is shown in the upper panel of Fig. \ref{fig:noise-comp1}. We also find that the difference in the noise levels is far lower than the difference in the noise-corrected power, indicating that the deadtime correction to the lightcurve is having an effect on the shape of the power spectra.
With a similar examination of a lower count rate observation where there is a similar proportional difference in the power calculated in the two cases, we find that there is a reversal in the relative sizes of the frequency dependent and the deadtime corrected lightcurve noise level, and that in this case the frequency-independent Poisson noise level without correction is closer to that calculated with Eq. \ref{eq:zhang-noise}. This is shown in Fig. \ref{fig:noise-comp2} for observation 4133010262.
This deviation is expected due to the different way in which we account for the deadtime recorded for events which are ultimately screened out by the analysis pipeline. In our approach in lower count rate observations the deadtime associated even with relatively low deadtime events (ie. those which do not meet the threshold to be considered a "very large event") directly affects the lightcurve and thus the power spectrum. Such events are not considered in the classical approach, as they will affect neither $r_{\text{pe}}$ nor $r_{\text{vle}}$. 

\section{Conclusions}
\label{sec:concs}
We have considered the effects of detector deadtime on nine observations of the X-ray Binary GX 339-4, examining the effects on an individual detector-by-detector basis and as a whole on the lightcurve and power spectrum.

We have found that at an individual detector level the overall effects of deadtime, in the deadtime of individual events detected and in the screening processes, are not significantly different between detectors. We find that the distributions of all event deadtimes are statistically distinct when considering any pair of observations, but that this distinction disappears when considering only X-ray events which are to form the lightcurves for individual observations. We find that for seven of the nine observations fitting the average deadtime per event across all the detectors with a constant value provides a very good quality of fit. We also find that when aggregated the proportional deadtime experienced by the individual FPMs is consistent with a constant value at the highest count rates. For lower count rate observations there can be individual detectors whose fractional deadtime is more than 4$\sigma$ or 5$\sigma$ from the mean, but these are anomalous results where detectors have been significantly affected by deadtime from events which are eventually screened by the processing pipeline.

Finally, we have considered the effects of deadtime on lightcurves and the power spectra derived from them. We find that the proportional detector deadtime in lightcurve bins can be adequately modelled by a linear fit dependent on observed count rate. This linear fit would imply that detector deadtime will likely not reach significant levels until such time as the count rate is at a point where telemetric deadtime will begin to affect the good time intervals for observations. This linear fit will breakdown at very high count rates, where the proportional deadtime accounts for a significant proportion of the time bins, as in such regimes the observed count rate might be significantly suppressed, but for count rates approaching tens of thousands of counts per second the effect of detector deadtime is well below $1\%$. The effects of deadtime on power spectra are not significant when comparing those derived from deadtime corrected lightcurves with those derived in the currently accepted manner which have a frequency-dependent noise subtraction. We find that there are deviations, both in the power spectra and in the noise level, when comparing power spectra from deadtime corrected lightcurves with those which do not have any adjustment to their lightcurves, and with frequency-dependent or independent noise subtractions. Given that there are no significant differences in the power spectra when using a deadtime corrected lightcurve, and with the increased demands computationally and in data storage, in utilising the unprocessed event files there does not appear to be a significant benefit in taking this approach. While it does account more fully for the effects of low-deadtime filtered events this doesn't appear to have a material effect on timing products. These results give us confidence that when going forward and using data from the \emph{NICER} X-ray Timing Instrument we can continue to process observations and obtain science products in the Fourier domain which can be comparable with those obtained historically from other instruments. Being able to compare such results allows for monitoring the longer term evolution of X-ray Binaries. Additionally, the emergence of other approaches, using machine learning \citep[e.g.][]{huppenkothen_accurate_2022} to infer the underlying shape of spectra show promise and may allow for an alternate approach to the issues of deadtime contamination of timing products.

\section*{Acknowledgements}

The authors thank the reviewers for their constructive support. All figures in this work were created with \texttt{Veusz} and \texttt{matplotlib}. This work is supported by the UKRI AIMLAC CDT, funded by grant EP/S023992/1.

%%%%%%%%%%%%%%%%%%%%%%%%%%%%%%%%%%%%%%%%%%%%%%%%%%
\section*{Data Availability}

This research has made use of publicly available data obtained through the High Energy Astrophysics Science Archive Research Center Online Service (https://heasarc.gsfc.nasa.gov/docs/archive.html), provided by the NASA/Goddard Space Flight Center. All analysis tools and products are available at: \url{https://github.com/robbie-webbe/DT_NICER}. For the purpose of open access, the author has applied a Creative Commons Attribution (CC BY) licence to any Author Accepted Manuscript version arising.



\bibliographystyle{rasti}
\bibliography{nicer_dt} % if your bibtex file is called example.bib

%\begin{thebibliography}{99}
%\end{thebibliography}

%%%%%%%%%%%%%%%%%%%%%%%%%%%%%%%%%%%%%%%%%%%%%%%%%%

%%%%%%%%%%%%%%%%% APPENDICES %%%%%%%%%%%%%%%%%%%%%

%\appendix



%%%%%%%%%%%%%%%%%%%%%%%%%%%%%%%%%%%%%%%%%%%%%%%%%%

% Don't change these lines
\bsp	% typesetting comment
\label{lastpage}
\end{document}